\documentclass[twocolumn,showpacs,preprintnumbers,superscriptaddress,amsmath,amssymb,nofootinbib]{revtex4}

\usepackage{graphicx}
\usepackage{bm}

\usepackage[normalem]{ulem}  
\usepackage[dvipdfmx]{color} 

\renewcommand\sout{\bgroup \color{red} \ULdepth=-.5ex \ULset}

\newcommand{\Slash}[1]{\ooalign{\hfil/\hfil\crcr$#1$}}
\newcommand{\Psfig}[2]{\includegraphics[width=#1]{#2}}
\newcommand{\PsfigII}[2]{\includegraphics[scale=#1]{#2}}

\newcommand{\SUN}[1]{\text{SU} ( #1 )}

\def\mev{\text{ MeV}}
\def\gev{\text{ GeV}}

\def\Kaellen{K\"{a}llen }

\def\trace{\text{tr}}

\begin{document}

\preprint{}

\title{Investigating the nature of light scalar mesons with
  semileptonic decays of $\bm{D}$ mesons}


\author{Takayasu~Sekihara} 
\email{sekihara@rcnp.osaka-u.ac.jp}
\affiliation{Research Center for Nuclear Physics
  (RCNP), Osaka University, Ibaraki, Osaka, 567-0047, Japan}

\author{Eulogio~Oset}
\email{oset@ific.uv.es}
\affiliation{ Departamento de F\'{\i}sica Te\'orica and IFIC, Centro
  Mixto Universidad de Valencia-CSIC, Institutos de Investigaci\'on de
  Paterna, Aptdo. 22085, 46071 Valencia, Spain }

\date{\today}

\begin{abstract}

  We study the semileptonic decays of $D_{s}^{+}$, $D^{+}$, and
  $D^{0}$ mesons into the light scalar mesons [$f_{0} (500)$,
  $K_{0}^{\ast} (800)$, $f_{0} (980)$, and $a_{0}(980)$] and the light
  vector mesons [$\rho (770)$, $\omega (782)$, $K^{\ast} (892)$, and
  $\phi (1020)$].  With the help of a chiral unitarity approach in
  coupled channels, we compute the branching fractions for scalar
  meson processes of the semileptonic $D$ decays in a simple way.
  Using current known values of the branching fractions, we make
  predictions for the branching fractions of the semileptonic decay
  modes with other scalar and vector mesons.  Furthermore, we
  calculate the $\pi ^{+} \pi ^{-}$, $\pi \eta$, $\pi K$, and $K^{+}
  K^{-}$ invariant mass distributions in the semileptonic decays of
  $D$ mesons, which will help us clarify the nature of the light
  scalar mesons.

\end{abstract}

\maketitle

\section{Introduction}

The recent experimental situation in hadron physics enables us to
utilize huge amounts of data on heavy hadrons, which contain charm or
bottom quark(s), for the investigation of hadron structures.
Especially, decay properties of heavy mesons can shed more light on
the nature of the light scalar mesons [$f_{0} (500)$, $K_{0}^{\ast}
(800)$, $f_{0} (980)$, and $a_{0}(980)$], which has been a hot topic
in hadron physics~\cite{Agashe:2014kda}.  For instance, the decay
$B_{s}^{0} \to J / \psi \pi ^{+} \pi ^{-}$ has been experimentally
measured in Refs.~\cite{Aaij:2011fx, Li:2011pg, Aaltonen:2011nk,
  Abazov:2011hv, LHCb:2012ae} for the study of the $f_{0} (500)$ and
$f_{0} (980)$ resonances, and they observed a pronounced peak for the
$f_{0} (980)$ while no evident signal was found for the $f_{0} (500)$.
Then a theoretical study~\cite{Liang:2014tia} followed the experiments
and reproduced ratios of experimental branching fractions at a
quantitative level, pointing out that $J / \psi + (s \bar{s})$
production in the $B_{s}^{0}$ decay and a hadronization of $s \bar{s}$
to $K \bar{K}$ are essential to understand the branching fractions of
the $B_{s}^{0}$ decay into $J / \psi f_{0} (980)$.  In the theoretical
study, the final state interaction between two pseudoscalar mesons is
calculated with the so-called chiral unitary
approach~\cite{Oller:1997ti, Kaiser:1998fi, Locher:1997gr,
  Oller:1997ng, Oller:1998hw, Oller:1998zr, Nieves:1999bx,
  Pelaez:2006nj, Albaladejo:2008qa}, in which the light scalar mesons
are obtained as dynamically generated resonances, and it is concluded
that the $f_{0} (980)$ has a substantial fraction of the strange
quarks.  The same hadronization scheme has been employed in
theoretical studies in Refs.~\cite{MartinezTorres:2009uk,
  Liang:2014tia, Xie:2014tma, Navarra:2015iea}.

\begin{figure}[!b]
  \centering
  \PsfigII{0.185}{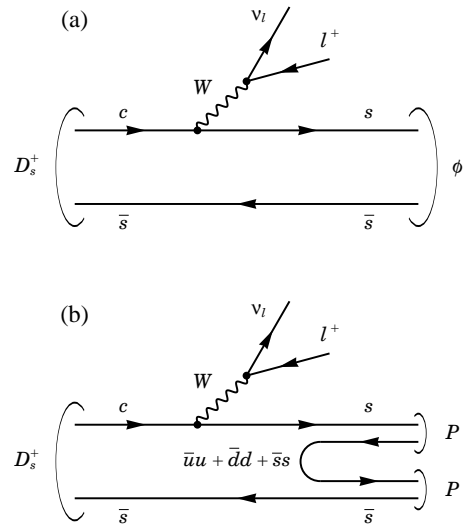}
  \caption{(a) Semileptonic decay of $D_{s}^{+}$ into $l^{+} \nu _{l}$
    and a primary $s \bar{s}$ pair.  (b) Semileptonic decay of
    $D_{s}^{+}$ into $l^{+} \nu _{l}$ and two pseudoscalar mesons $P$
    with a hadronization.}
\label{fig:12}
\end{figure}

In this paper, we consider the semileptonic decay of $D \to
\text{hadron(s)} + l^{+} \nu _{l}$, extending a discussion for the
semileptonic $B$ decays into $D_{s 0}^{\ast} (2317)$ and $D_{0}^{\ast}
(2400)$ resonances in Ref.~\cite{Navarra:2015iea}.  The semileptonic
$D$ decays have been experimentally investigated in, {\it e.g.},
BES~\cite{Ablikim:2006ah, Ablikim:2006hw}, FOCUS~\cite{Link:2004gp,
  Link:2005xe}, BaBar~\cite{Aubert:2008rs, delAmoSanchez:2010fd}, and
CLEO~\cite{Yelton:2009aa, Ecklund:2009aa, Martin:2011rd,
  Yelton:2010js, CLEO:2011ab}.  Here, in order to grasp how the
semileptonic decay takes place, let us consider the $D_{s}^{+}$ meson.
Since the constituent quark component of $D_{s}^{+}$ is $c \bar{s}$,
we expect a Cabibbo favored semileptonic decay of $c \to s \, l^{+} \,
\nu _{l}$ and hence the decay $D_{s}^{+} \to ( s \bar{s} ) \, l^{+} \,
\nu _{l}$ with $s \bar{s}$ being the vector meson $\phi (1020)$, which
is depicted in Fig.~\ref{fig:12}(a).  Actually this semileptonic decay
mode has been observed in experiments, and its branching fraction to
the total decay width is $\mathcal{B}[D_{s}^{+} \to \phi (1020) \,
e^{+} \, \nu _{e}] = 2.49 \pm 0.14 \%$~\cite{Agashe:2014kda} (see
Table~\ref{tab:Br}, in which we list branching fractions for the
semileptonic decays of $D_{s}^{+}$, $D^{+}$, and $D^{0}$ reported by
the Particle Data Group).  On the other hand, we cannot
straightforwardly extend the discussion to the scalar meson
productions in the final state of the semileptonic decays, since the
structure of the scalar mesons, whether $q \bar{q}$ or some exotic
one, is still controversial.  In this study we consider the production
of the $f_{0} (980)$ or $f_{0} (500)$ as dynamically generated
resonances in the semileptonic $D_{s}^{+}$ decay, so we have to
introduce an extra $\bar{q} q$ pair to make a hadronization as shown
in Fig.~\ref{fig:12}(b).  The introduction of an extra $\bar{q} q$
pair to make a hadronization has been performed in
Refs.~\cite{MartinezTorres:2009uk, Liang:2014tia, Xie:2014tma,
  Navarra:2015iea}.  In this study we apply the same method of the
hadronization to the semileptonic decays of $D$ mesons so as to
investigate the nature of the light scalar mesons.

\def\arraystretch{1.25}
\begin{table}
  \caption{Branching fractions for the semileptonic decays of
    $D_{s}^{+}$, $D^{+}$, and $D^{0}$ reported by the Particle Data
    Group~\cite{Agashe:2014kda}.  In this Table we only show decay
    modes relevant to this study.}
 \label{tab:Br}
  \begin{ruledtabular}
    \begin{tabular*}{8.6cm}{@{\extracolsep{\fill}}lc}
      \multicolumn{2}{c}{$D_{s}^{+}$} \\
      \hline
      Mean life [s] & $(500 \pm 7) \times 10^{-15}$ \\
      $\mathcal{B}[\phi (1020) e^{+} \nu _{e}]$ & 
      $(2.49 \pm 0.14) \times 10^{-2}$ \\
      $\mathcal{B}[\omega (782) e^{+} \nu _{e}]$ & 
      $< 2.0 \times 10^{-3}$ \\
      $\mathcal{B}[K^{\ast} (892)^{0} e^{+} \nu _{e}]$ & 
      $(1.8 \pm 0.7) \times 10^{-3}$ \\
      $\mathcal{B}[f_{0} (980) e^{+} \nu _{e}, \, 
      f_{0} (980) \to \pi ^{+} \pi ^{-}]$ & 
      $(2.00 \pm 0.32) \times 10^{-3}$ \\
      \\
      \multicolumn{2}{c}{$D^{+}$} \\
      \hline
      Mean life [s] & $(1040 \pm 7) \times 10^{-15}$ \\
      $\mathcal{B}[\bar{K}^{\ast} (892)^{0} e^{+} \nu _{e}, \, 
      \bar{K}^{\ast} (892)^{0} \to K^{-} \pi ^{+}]$ & 
      $(3.68 \pm 0.10) \times 10^{-2}$ \\
      $\mathcal{B}[(K^{-} \pi ^{+})_{s\text{-wave}} e^{+} \nu _{e}]$ & 
      $(2.32 \pm 0.10) \times 10^{-3}$ \\
      $\mathcal{B}[\bar{K}^{\ast} (892)^{0} \mu ^{+} \nu _{\mu}, \, 
      \bar{K}^{\ast} (892)^{0} \to K^{-} \pi ^{+}]$ & 
      $(3.52 \pm 0.10) \times 10^{-2}$ \\
      $\mathcal{B}[\rho (770)^{0} e^{+} \nu _{e}]$ & 
      $(2.18 ^{+0.17}_{-0.25}) \times 10^{-3}$ \\
      $\mathcal{B}[\rho (770)^{0} \mu ^{+} \nu _{\mu}]$ & 
      $(2.4 \pm 0.4) \times 10^{-3}$ \\
      $\mathcal{B}[\omega (782) e^{+} \nu _{e}]$ & 
      $(1.82 \pm 0.19) \times 10^{-3}$ \\
      $\mathcal{B}[\phi (1020) e^{+} \nu _{e}]$ & 
      $< 9 \times 10^{-5}$ \\
      \\
      \multicolumn{2}{c}{$D^{0}$} \\
      \hline
      Mean life [s] & $(410.1 \pm 1.5) \times 10^{-15}$ \\
      $\mathcal{B}[K^{\ast} (892)^{-} e^{+} \nu _{e}]$ & 
      $(2.16 \pm 0.16) \times 10^{-2}$ \\
      $\mathcal{B}[K^{\ast} (892)^{-} \mu ^{+} \nu _{\mu}]$ & 
      $(1.90 \pm 0.24) \times 10^{-2}$ \\
      $\mathcal{B}[K^{-} \pi ^{0} e^{+} \nu _{e}]$ & 
      $(1.6 ^{+1.3}_{-0.5}) \times 10^{-2}$ \\
      $\mathcal{B}[\bar{K}^{0} \pi ^{-} e^{+} \nu _{e}]$ & 
      $(2.7 ^{+0.9}_{-0.7}) \times 10^{-2}$ \\
      $\mathcal{B}[\rho (770)^{-} e^{+} \nu _{e}]$ & 
      $(1.9 \pm 0.4) \times 10^{-3}$ \\
    \end{tabular*}
  \end{ruledtabular}
\end{table}
\def\arraystretch{1.0}

Utilizing the semileptonic decay of a heavy hadron provides us with
two advantages when we investigate the internal structure of hadrons
in the final state of the semileptonic decay.  First, Cabibbo favored
and suppressed processes enable us to specify flavors of quarks
contained in final state hadrons.  Second, the semileptonic decay of
the heavy hadron to two light hadrons $+ l^{+} \nu _{l}$ brings a
suitable condition to measure effects of the final state interaction
of the two light hadrons, since the leptons and hadrons in the final
state interact with each other only weakly.  

Theoretical work on the issues of the semileptonic $D$ decays is
already available.  In Ref.~\cite{Bediaga:2003zh}, using QCD sum
rules, the $D_{s}^{+}$ and $D^{+}$ semileptonic decays into $f_{0}
(980)$ are considered concluding that the importance of up and down
quarks in the $f_{0} (980)$ is not negligible.  In
Ref.~\cite{Ke:2009ed} the $D_{s}^{+} \to f_{0} (980) e^{+} \nu _{e}$
reaction is analyzed from the point of view of the $f_{0} (980)$ being
a $q \bar{q}$ state, concluding that $s \bar{s}$ component of the
$f_{0} (980)$ may not be dominant.  In Ref.~\cite{Achasov:2012kk} the
$D_{s}^{+} \to \pi ^{+} \pi ^{-} e^{+} \nu _{e}$ reaction is studied
concluding that it supports the dominant four quark nature of the
$f_{0} (500)$ and $f_{0} (980)$.  Similar conclusions about the four
quark nature of the scalar mesons are reached in the work
of~\cite{Fariborz:2011xb, Fariborz:2014mpa}.  Research along the same
line is done in Ref.~\cite{Wang:2009azc}, looking for likely
reasonable ratios that would help distinguish between the two and four
quark structure of the scalar mesons.  

Another line of research is done using light-cone sum rules to
evaluate the form factors appearing in the
process~\cite{Meissner:2013hya}.  This line of research is applied in
many related processes, rare decays like $B_{s} \to \pi ^{+} \pi ^{-}
l^{+} l^{-}$ in~\cite{Wang:2015paa}, $B_{s} \to K^{( \ast )} l
\bar{\nu}$ in~\cite{Meissner:2013pba}, $B^{0}_{(s)}\to J/\psi \pi ^{+}
\pi ^{-}$ and $B_{s} \to \pi ^{+} \pi ^{-} \mu ^{+} \mu ^{-}$ decays
in~\cite{Wang:2015uea}, or semileptonic decays~\cite{Kang:2013jaa,
  Doring:2013wka, Shi:2015kha}. In some cases the meson final state
interaction is further implemented using the Omnes
representation~\cite{Meissner:2013hya, Doring:2013wka}, while in other
cases Breit--Wigner or Flatte structures are implemented and
parametrized to account for the resonances observed in the experiment.

In contrast to these pictures, in the present study we treat the
scalar mesons as dynamically generated resonances from two
pseudoscalar mesons in the so-called chiral unitary approach.  Then we
describe the semileptonic decays of $D$ mesons in an economical way
for hadronization as done in Refs.~\cite{MartinezTorres:2009uk,
  Liang:2014tia, Xie:2014tma, Navarra:2015iea}.

This paper is organized as follows. In Sec.~\ref{sec:form} we
formulate the semileptonic decay widths of $D_{s}^{+}$, $D^{+}$, and
$D^{0}$ into the light scalar and vector mesons and give our model of
the hadronization.  We also calculate meson--meson scattering
amplitudes to generate dynamically the scalar mesons.  In
Sec.~\ref{sec:results} we show our numerical results of the
semileptonic decay widths of $D_{s}^{+}$, $D^{+}$, and $D^{0}$.  We
predict branching fractions which are not reported by the Particle
Data Group and show invariant mass distributions of the two
pseudoscalar mesons from the scalar and vector mesons.
Section~\ref{sec:conclusion} is devoted to drawing the conclusion of
this study.

\section{Formulation}
\label{sec:form}

\begin{table}
  \caption{Semileptonic decay modes of $D_{s}^{+}$, $D^{+}$, and
    $D^{0}$ considered in this study.  The lepton flavor $l$ is $e$
    and $\mu$.  We also specify Cabibbo favored/suppressed process for
    each decay mode; the semileptonic decay into two pseudoscalar
    mesons is judged with the discussions given in
    Sec.~\ref{sec:had}. }
 \label{tab:mode}
  \begin{ruledtabular}
    \begin{tabular*}{8.6cm}{@{\extracolsep{\fill}}lc}
      \multicolumn{2}{c}{$D_{s}^{+}$} 
      \\
      \hline
      $\phi (1020) \, l^{+} \, \nu _{l}$ & favored 
      \\
      $K^{\ast} (892)^{0} \, l^{+} \, \nu _{l}$ & suppressed 
      \\
      $\pi ^{+} \pi ^{-} \, l^{+} \, \nu _{l}$ & favored
      \\
      $K^{+} K^{-} \, l^{+} \, \nu _{l}$ & favored
      \\
      $\pi ^{-} K^{+} \, l^{+} \, \nu _{l}$ & suppressed
      \\
      \\
      \multicolumn{2}{c}{$D^{+}$} 
      \\
      \hline
      $\bar{K}^{\ast} (892)^{0} \, l^{+} \, \nu _{l}$ & favored
      \\
      $\rho (770)^{0} \, l^{+} \, \nu _{l}$ & suppressed
      \\
      $\omega (782) \, l^{+} \, \nu _{l}$ & suppressed
      \\
      $\pi ^{+} \pi ^{-} \, l^{+} \, \nu _{l}$ & suppressed
      \\
      $\pi ^{0} \eta \, l^{+} \, \nu _{l}$ & suppressed
      \\
      $K^{+} K^{-} \, l^{+} \, \nu _{l}$ & suppressed
      \\
      $\pi ^{+} K^{-} \, l^{+} \, \nu _{l}$ & favored
      \\
      \\
      \multicolumn{2}{c}{$D^{0}$} 
      \\
      \hline
      $K^{\ast} (892)^{-} \, l^{+} \, \nu _{l}$ & favored
      \\
      $\rho (770)^{-} \, l^{+} \, \nu _{l}$ & suppressed
      \\
      $\pi ^{-} \eta \, l^{+} \, \nu _{l}$ & suppressed
      \\
      $K^{0} K^{-} \, l^{+} \, \nu _{l}$ & suppressed
      \\
      $\pi ^{-} \bar{K}^{0} \, l^{+} \, \nu _{l}$ & favored
      \\
    \end{tabular*}
  \end{ruledtabular}
\end{table}

In this section we formulate the semileptonic decay widths of
$D_{s}^{+}$, $D^{+}$, and $D^{0}$ into light scalar and vector mesons:
\begin{equation}
D_{s}^{+} , \, D^{+} , \, D^{0} \to 
\begin{cases}
S l^{+} \nu _{l} , \quad S \to P P , \\
V l^{+} \nu _{l} , 
\end{cases}
\end{equation}
where $S$, $V$, and $P$ represent the light scalar, vector, and
pseudoscalar mesons, respectively, and the lepton flavor $l$ can be
$e$ and $\mu$.  Explicit decay modes are listed in
Table~\ref{tab:mode}.  In order to formulate the decay width, we
consider first the semileptonic decay amplitudes and widths in
Section~\ref{sec:decay} and next hadronizations into scalar and vector
mesons in Section~\ref{sec:had}.  Scattering amplitudes of two
pseudoscalar mesons are then constructed in the chiral unitary
approach for the description of the scalar mesons in
Section~\ref{sec:ChUA}.  Throughout this study we assume isospin
symmetry for light hadrons.

\subsection{Amplitudes and widths of semileptonic $\bm{D}$ decays}
\label{sec:decay}

In general, we can express the decay amplitude of $D \to
\text{hadron(s)} + l^{+} \nu _{l}$, $T_{D}$, by using the propagator
of the $W$ boson and its couplings to leptons and quarks, which can be
replaced with the Fermi coupling constant $G_{\rm F}$.  At this stage
we do not fix the number of the final state hadrons.  In a similar
manner to the formulation in Ref.~\cite{Navarra:2015iea}, the explicit
form of $T_{D}$ becomes
\begin{equation}
T_{D} = - i \frac{G_{\rm F}}{\sqrt{2}} 
L^{\alpha} Q_{\alpha} \times V_{\rm had} .
\label{eq:TD}
\end{equation}
The factor $V_{\rm had}$ consists of the wave function of quarks
inside the $D$ meson, the hadronization contribution in the final
state, and the Cabibbo--Kobayashi--Maskawa matrix element for the
transition from the charm to a light quark.  The explicit form of
$V_{\rm had}$ will be determined in the next subsection.  The lepton
and quark parts of the $W$ boson couplings are defined as.
\begin{equation}
L^{\alpha} \equiv 
\overline{u}_{\nu} \gamma ^{\alpha} ( 1 - \gamma _{5} ) v_{l} ,
\quad 
Q_{\alpha} \equiv 
\overline{u}_{q} \gamma _{\alpha} ( 1 - \gamma _{5} ) u_{c} ,
\label{eq:LQ}
\end{equation}
respectively, where $u_{\nu}$, $v_{l}$, $u_{q}$, and $u_{c}$ are the
Dirac spinors corresponding to the neutrino, lepton $l^{+}$, light
quark $q$, and charm quark, respectively.

Let us now calculate the squared amplitude for the semileptonic $D$
decay widths, in which we average (sum) the polarizations of the
initial-state quarks (final state leptons and quarks).  Therefore, in
terms of the amplitude in Eq.~\eqref{eq:TD}, we can obtain the squared
decay amplitude as
\begin{equation}
\frac{1}{2} \sum _{\rm pol} | T_{D} |^{2}
= \frac{| G_{\rm F} V_{\rm had} |^{2}}{4}
\sum _{\rm pol} | L^{\alpha} Q_{\alpha} |^{2}
\end{equation}
where the factor $1/2$ comes from the average of the charm quark
polarization in the initial state.  We can further calculate the lepton
and quark parts in the amplitude~\eqref{eq:LQ}, by using the
conventions of the Dirac spinors and traces of Dirac $\gamma$ matrices
summarized in Appendix~\ref{app:1}, which lead to
\begin{align}
\sum _{\rm pol} L^{\alpha} L^{\dagger \beta}
= & \trace \left [ \gamma ^{\alpha} ( 1 - \gamma _{5} )
\frac{\Slash{p}_{l} - m_{l}}{2 m_{l}}
( 1 + \gamma _{5} ) \gamma ^{\beta} 
\frac{\Slash{p}_{\nu} + m_{\nu}}{2 m_{\nu}}
 \right ]
\notag \\
= & 2 \frac{p_{l}^{\alpha} p_{\nu}^{\beta} + p_{\nu}^{\alpha} p_{l}^{\beta}
- p_{l} \cdot p_{\nu} g^{\alpha \beta} + i \epsilon ^{\alpha \beta \rho \sigma}
p_{l \rho} p_{\nu \sigma}}{m_{l} m_{\nu}} ,
\end{align}
where $p_{l}$ and $p_{\nu}$ ($m_{l}$ and $m_{\nu}$) are momenta
(masses) of the lepton $l^{+}$ and neutrino, respectively, and
\begin{align}
\sum _{\rm pol} Q_{\alpha} Q_{\beta}^{\dagger}
= & \trace \left [ \gamma _{\alpha} ( 1 - \gamma _{5} )
\frac{\Slash{p}_{c} + m_{c}}{2 m_{c}}
( 1 + \gamma _{5} ) \gamma _{\beta} 
\frac{\Slash{p}_{q} + m_{q}}{2 m_{q}}
 \right ]
\notag \\
= & 2 \frac{p_{c \alpha} p_{q \beta} + p_{q \alpha} p_{c \beta}
- p_{c} \cdot p_{q} g_{\alpha \beta} + i \epsilon _{\alpha \beta \rho \sigma}
p_{c}^{\rho} p_{q}^{\sigma}}{m_{c} m_{q}} ,
\end{align}
with the momenta (masses) of the charm and light quarks, $p_{c}$ and
$p_{q}$ ($m_{c}$ and $m_{q}$), respectively.\footnote{The momentum
  $p_{q}$ is for a quark in the primary $q \bar{q}$ pair after the $W$
  boson emission, which means that the momentum $p_{q}$ is carried by
  the constituent quark.  Accordingly, $m_{q}$ is the mass of the
  constituent quark rather than of the current quark.  In this sense,
  $m_{q}$ respects the flavor $\SUN{3}$ symmetry.}  Then with a
straightforward calculation we have
\begin{equation}
\sum _{\rm pol} | L^{\alpha} Q_{\alpha} |^{2}
= \frac{16 ( p_{l} \cdot p_{c} ) ( p_{\nu} \cdot p_{q} )}
{m_{l} m_{\nu} m_{c} m_{q}} .
\end{equation}
Now let us rewrite the momenta of quarks by using those of hadrons in
the following manner:
\begin{equation}
\frac{p_{c}^{\mu}}{m_{c}} = \frac{p_{D}^{\mu}}{m_{D}} ,
\quad 
\frac{p_{q}^{\mu}}{m_{q}} = \frac{p_{R}^{\mu}}{m_{R}} ,
\end{equation}
where we have neglected the relative internal momenta of the quarks,
which are typically small compared to the masses of quarks.  Here
$m_{D}$ and $m_{R}$ ($p_{D}$ and $p_{R}$) are the masses (momenta) of
the $D$ and $R = S$, $V$ mesons, respectively.  With these
translations the square of $L^{\alpha} Q_{\alpha}$ with polarization
summation becomes
\begin{equation}
\sum _{\rm pol} | L^{\alpha} Q_{\alpha} |^{2}
= \frac{16 ( p_{l} \cdot p_{D} ) ( p_{\nu} \cdot p_{R} )}
{m_{l} m_{\nu} m_{D} m_{R}} .
\end{equation}
Therefore, we obtain the squared decay amplitude as:
\begin{equation}
\frac{1}{2} \sum _{\rm pol} | T_{D} |^{2}
= \frac{4 | G_{\rm F} V_{\rm had} |^{2}}{m_{l} m_{\nu} m_{D} m_{R}}
( p_{l} \cdot p_{D} ) ( p_{\nu} \cdot p_{R} ) .
\label{eq:amp2}
\end{equation}

With the above squared amplitude we can compute the decay width. We
will be interested in two types of decays: three-body decays for
vector mesons such as $D_{s}^{+} \to \phi (1020) \, e^{+} \, \nu
_{e}$, and four-body decays for scalar mesons constructed from two
pseudoscalar mesons such as $D_{s}^{+} \to \pi ^{+} \pi ^{-} \, e^{+}
\, \nu _{e}$.  As it will be seen, both decay types can be described
by the amplitude $T_{D}$ with different assumptions for $V_{\rm had}$:
$V_{\rm had}^{(v)}$ and $V_{\rm had}^{(s)}$ respectively.  

The formula for the three-body decay is given
by~\cite{Agashe:2014kda}:
\begin{equation}
\Gamma _{3} = \frac{m_{l} m_{\nu}}{128 \pi ^{5} m_{D}^{2}}
\int d M_{\rm inv}^{( l \nu )} 
P_{\rm cm} \tilde{p}_{\nu} \int d \Omega \int d \tilde{\Omega}_{\nu}
\frac{1}{2} \sum _{\rm pol} | T_{D} |^{2} ,
\end{equation}
where $P_{\rm cm}$ is the momentum of the final state vector meson in
the $D$ rest frame and $\tilde{p}_{\nu}$ is the momentum of the
neutrino in the $l \nu$ rest frame, both of which are evaluated as
\begin{equation}
P_{\rm cm} = \frac{\lambda ^{1/2} (m_{D}^{2}, \, 
[M_{\rm inv}^{( l \nu )}]^{2}, \, m_{V}^{2}) }{2 m_{D}} , 
\end{equation}
\begin{equation}
\tilde{p}_{\nu} = \frac{\lambda ^{1/2} ([M_{\rm inv}^{( l \nu )}]^{2}, 
\, m_{l}^{2}, \, m_{\nu}^{2}) }{2 M_{\rm inv}^{( l \nu )}} ,
\label{eq:pnu_tilde}
\end{equation}
with the \Kaellen function $\lambda (x, \, y, \, z) = x^{2} + y^{2} +
z^{2} - 2 x y - 2 y z - 2 z x$ and the vector meson mass $m_{V}$.  The
tilde on characters for leptons indicates that they are evaluated in
the $l \nu$ rest frame.  The solid angles $\Omega$ and $\tilde{\Omega}
_{\nu}$ are for the vector meson in the $D$ rest frame and for the
neutrino in the $l \nu$ rest frame, respectively, and $M_{\rm inv}^{(l
  \nu )}$ is the $l \nu$ invariant mass.  The integral range of
$M_{\rm inv}^{( l \nu )}$ is $[m_{l} + m_{\nu} , \, m_{D} - m_{V} ]$.
Substituting the squared amplitude with that in Eq.~\eqref{eq:amp2},
we obtain
\begin{align}
\Gamma _{3} = & \frac{| G_{\rm F} |^{2}}
{32 \pi ^{5} m_{D}^{3} m_{V}}
\int d M_{\rm inv}^{( l \nu )} 
P_{\rm cm} \tilde{p}_{\nu} \int d \Omega \int d \tilde{\Omega}_{\nu}
\notag \\
& \times \left | V_{\rm had}^{(v)} \right | ^{2} 
( p_{l} \cdot p_{D} ) ( p_{\nu} \cdot p_{V} ) .
\end{align}
In general, the hadronization part $V_{\rm had}$ may depend on the
energy and scattering angles, and hence one cannot put it out of the
integral.  In this study, however, $V_{\rm had}^{(v)}$ will be simply
constructed, so that this will not depend on $M_{\rm inv}^{( l \nu )}$
nor the angle, as we will see in the next subsection.  Furthermore,
the integral of the solid angle $\tilde{\Omega}_{\nu}$ is performed in
the $l \nu$ rest frame as~\cite{Navarra:2015iea}
\begin{align}
& \int d \tilde{\Omega}_{\nu} 
( p_{l} \cdot p_{D} ) ( p_{\nu} \cdot p_{V} ) 
\notag \\
& = \int d \tilde{\Omega}_{\nu}
( \tilde{E}_{l} \tilde{E}_{D} + \tilde{\bm{p}}_{\nu} \cdot \tilde{\bm{p}}_{D} )
( \tilde{E}_{\nu} \tilde{E}_{V} - \tilde{\bm{p}}_{\nu} \cdot \tilde{\bm{p}}_{D} )
\notag \\
& = 4 \pi \tilde{E}_{l} \tilde{E}_{\nu} \tilde{E}_{D} \tilde{E}_{V} 
- \frac{4 \pi}{3} | \tilde{\bm{p}}_{\nu} |^{2} | \tilde{\bm{p}}_{D} |^{2} 
\notag \\
& = \pi [ M_{\rm inv}^{(l \nu )} ]^{2} 
\left ( \tilde{E}_{D} \tilde{E}_{V} - \frac{1}{3} | \tilde{\bm{p}}_{D} |^{2} 
\right ) ,
\end{align}
where $\tilde{E}$ and $\tilde{\bm{p}}$ are the energies and momenta in
the $l \nu$ rest frame.  At the first equality we have used relations
$\tilde{\bm{p}}_{l} = - \tilde{\bm{p}}_{\nu}$ and $\tilde{\bm{p}}_{V}
= \tilde{\bm{p}}_{D}$, while at the third equality we have used
relations obtained by neglecting masses of leptons:
\begin{equation}
\tilde{E}_{l} = \tilde{E}_{\nu} =
| \tilde{\bm{p}}_{\nu} | = \frac{M_{\rm inv}^{(l \nu )}}{2} .
\end{equation}
The energies and momentum of hadrons in the $l \nu$ rest frame can be
exactly evaluated as
\begin{equation}
\tilde{E}_{D} = \frac{m_{D}^{2} + [ M_{\rm inv}^{( l \nu )}]^{2} 
- m_{V}^{2}}{2 M_{\rm inv}^{( l \nu )}} ,
\end{equation}
\begin{equation}
\tilde{E}_{V} = \frac{m_{D}^{2} - [ M_{\rm inv}^{( l \nu )}]^{2} 
- m_{V}^{2}}{2 M_{\rm inv}^{( l \nu )}} ,
\end{equation}
and $| \tilde{\bm{p}}_{D} |^{2} = \tilde{E}_{D}^{2} - m_{D}^{2}$.
As a consequence, we have
\begin{align}
\Gamma _{3} = \frac{\left | G_{\rm F} V_{\rm had}^{(v)} \right | ^{2}}
{8 \pi ^{3} m_{D}^{3} m_{V}}
& \int d M_{\rm inv}^{( l \nu )} 
P_{\rm cm} \tilde{p}_{\nu} [ M_{\rm inv}^{(l \nu )} ]^{2} 
\notag \\
& \times 
\left ( \tilde{E}_{D} \tilde{E}_{V} - \frac{1}{3} | \tilde{\bm{p}}_{D} |^{2} 
\right ) ,
\end{align}
where we have performed the integral of the solid angle $\Omega$.

In a similar way, we can evaluate the decay width for the four-body
final state.  The formula for the four-body decay is given by
\begin{align}
\Gamma _{4} = & \frac{m_{l} m_{\nu}}{2048 \pi ^{8} m_{D}^{2}}
\int d M_{\rm inv}^{(h h)} \int d M_{\rm inv}^{( l \nu )} 
P_{\rm cm}^{\prime} \tilde{p}_{h} \tilde{p}_{\nu}
\notag \\ & \times 
\int d \Omega ^{\prime} \int d \tilde{\Omega}_{h} \int d \tilde{\Omega}_{\nu} 
\frac{1}{2} \sum _{\rm pol} | T_{D} |^{2} ,
\end{align}
where $M_{\rm inv}^{(h h)}$ is the invariant mass of the two-meson
system ($h h$), $P_{\rm cm}^{\prime}$ is the center-of-mass momentum
of the two-meson system in the $D$ rest frame and $\tilde{p}_{h}$ is
the momentum of a meson in the $h h$ rest frame, both of which are
evaluated as
\begin{equation}
P_{\rm cm}^{\prime} = \frac{\lambda ^{1/2} (m_{D}^{2} , \, 
[M_{\rm inv}^{(h h)}]^{2} , \, [M_{\rm inv}^{( l \nu )}]^{2})}{2 m_{D}} ,
\end{equation}
\begin{equation}
\tilde{p}_{h} = \frac{\lambda ^{1/2} ([M_{\rm inv}^{(h h)}]^{2}, \, 
m_{h}^{2}, \, m_{h}^{\prime 2}) }{2 M_{\rm inv}^{(h h)}} ,
\end{equation}
with the meson masses $m_{h}$ and $m_{h}^{\prime}$.  The momentum of
the neutrino in the $l \nu$ rest frame $\tilde{p}_{\nu}$ is given in
Eq.~\eqref{eq:pnu_tilde}.  The solid angles $\Omega ^{\prime}$ and
$\tilde{\Omega} _{h}$ are for the two-meson system in the $D$ rest
frame and for a meson in the $h h$ rest frame, respectively.  The
tilde on characters for mesons indicates that they are evaluated in
the $h h$ rest frame.  Since we are interested in the meson--meson
invariant mass distributions for the semileptonic $D$ decay, we
calculate the differential decay width $d \Gamma _{4} / d M_{\rm
  inv}^{(h h)}$.  Then in a similar manner to the case of the
three-body decay, we have
\begin{align}
\frac{d \Gamma _{4}}{d M_{\rm inv}^{(h h)}} 
= \frac{\left | G_{\rm F} V_{\rm had}^{(s)} \right | ^{2}}
{32 \pi ^{5} m_{D}^{3} M_{\rm inv}^{(h h)}}
& \int d M_{\rm inv}^{( l \nu )} 
P_{\rm cm}^{\prime} \tilde{p}_{h} \tilde{p}_{\nu} [M_{\rm inv}^{( l \nu )}]^{2}
\notag \\
& \times 
\left ( \tilde{E}_{D} \tilde{E}_{S} - \frac{1}{3} | \tilde{\bm{p}}_{D} |^{2} 
\right ) ,
\end{align}
where we have performed the integrals with respect to the solid angles
$\Omega ^{\prime}$ and $\tilde{\Omega} _{h}$.  We mention that $V_{\rm
  had}^{(s)}$ will be simply constructed as well, so that this can be
put out of the integral, as we will see in the next subsection.  The
two-meson invariant mass $M_{\rm inv}^{(h h)}$ can take a value within
$[m_{h} + m_{h}^{\prime}, \, m_{D} - m_{l} - m_{\nu}]$, while the
integral range of $M_{\rm inv}^{( l \nu )}$ is $[m_{l} + m_{\nu} , \,
m_{D} - M_{\rm inv}^{(h h)} ]$.  The energies and momentum of hadrons
in the parentheses can be exactly evaluated as
\begin{equation}
\tilde{E}_{D} = \frac{m_{D}^{2} + [ M_{\rm inv}^{( l \nu )}]^{2} 
- [ M_{\rm inv}^{(h h)} ]^{2}}{2 M_{\rm inv}^{( l \nu )}} ,
\end{equation}
\begin{equation}
\tilde{E}_{S} = \frac{m_{D}^{2} - [ M_{\rm inv}^{( l \nu )}]^{2} 
- [ M_{\rm inv}^{(h h)} ]^{2}}{2 M_{\rm inv}^{( l \nu )}} ,
\end{equation}
and $| \tilde{\bm{p}}_{D} |^{2} = \tilde{E}_{D}^{2} - m_{D}^{2}$.

\subsection{Hadronizations}
\label{sec:had}

\begin{figure}[!b]
  \centering
  \PsfigII{0.185}{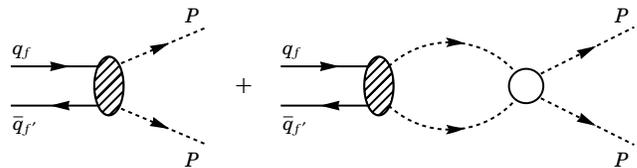}
  \caption{Diagrammatic representation of the direct plus
    rescattering processes for two pseudoscalar mesons.  The solid and
    dashed lines denote quarks and pseudoscalar mesons, respectively.
    The shaded ellipses indicate the hadronization of a
    quark--antiquark pair into two pseudoscalar mesons, while the open
    circle indicates the rescattering of two pseudoscalar mesons.}
\label{fig:3}
\end{figure}

Next we fix the mechanism for the appearance of the scalar and vector
mesons in the final state of the semileptonic decay.  We here note
that, for the scalar and vector mesons in the final state, the
hadronization processes should be different from each other according
to their structure.  For the scalar mesons, we employ the chiral
unitary approach~\cite{Oller:1997ti, Kaiser:1998fi, Locher:1997gr,
  Oller:1997ng, Oller:1998hw, Oller:1998zr, Nieves:1999bx,
  Pelaez:2006nj, Albaladejo:2008qa}, in which the scalar mesons are
dynamically generated from the interaction of two pseudoscalar mesons
governed by the chiral Lagrangians.  Therefore, in this picture the
light quark--antiquark pair after the $W$ boson emission gets
hadronized by adding an extra $\bar{q} q$ with the quantum number of
the vacuum, $\bar{u} u + \bar{d} d + \bar{s} s$, which results in two
pseudoscalar mesons in the final state [see Fig.~\ref{fig:12}(b)].
Then the scalar mesons are obtained as a consequence of the final
state interaction of the two pseudoscalar mesons as diagrammatically
shown in Fig.~\ref{fig:3}.  For the vector mesons, on the other hand,
hadronization with an extra $\bar{q} q$ is unnecessary since they are
expected to consist genuinely of a light quark--antiquark pair [see
Fig.~\ref{fig:12}(a)].

\subsubsection{Scalar mesons}

First we consider processes with the scalar mesons in the final state
as the dynamically generated resonances.  The basic idea of the
hadronization with an extra $\bar{q} q$ with the quantum number of the
vacuum has already shown in Refs.~\cite{MartinezTorres:2009uk,
  Liang:2014tia, Xie:2014tma, Navarra:2015iea}.  We start with the $q
\bar{q}$ matrix $M$:
\begin{equation}
  M = \left(
    \begin{array}{ccc}
      u \bar{u} & u \bar{d}  & u \bar{s} \\
      d \bar{u} & d \bar{d}  & d \bar{s} \\
      s \bar{u} & s \bar{d}  & s \bar{s} \\
    \end{array}
  \right) .
\end{equation}
One can easily check that this matrix has the property
\begin{equation}
  M \cdot M = M ( \bar{u} u + \bar{d} d + \bar{s} s ) .
\end{equation}
With this property, a $q_{f} \bar{q}_{f^{\prime}}$ pair after the $W$
boson emission can be added by an extra $\bar{q} q$ to be
\begin{equation}
  q_{f} \bar{q}_{f^{\prime}} \to ( M \cdot M )_{f f^{\prime}} ,
\end{equation}
where $f$ denotes the flavor of light quarks: $q_{1} = u$, $q_{2} =
d$, and $q_{3} = s$.  Next we rewrite the matrix $M$ in terms of the
matrix $\phi$ for pseudoscalar mesons
\begin{widetext}
\begin{equation}
\phi=
\left(
  \begin{array}{ccc}
    \frac{1}{\sqrt{2}} \pi ^{0} + \frac{1}{\sqrt{3}} \eta 
    + \frac{1}{\sqrt{6}} \eta ^{\prime}
    & \pi^{+}  & K^{+} \\
    \pi ^{-} &  - \frac{1}{\sqrt{2}} \pi ^{0} 
    + \frac{1}{\sqrt{3}} \eta + \frac{1}{\sqrt{6}} \eta ^{\prime} & K^{0} \\
    K^{-} & \bar{K}^{0} &  - \frac{1}{\sqrt{3}} \eta 
    + \sqrt{\frac{2}{3}} \eta ^{\prime} \\
  \end{array}
\right),
\end{equation}
\end{widetext}
where we have taken into account the $\eta$--$\eta ^{\prime}$ mixing
in a standard way~\cite{Bramon:1992kr}.  In this scheme we can
calculate the weight of each pair of pseudoscalar mesons in the
hadronization.  Namely, the $s \bar{s}$ pair gets hadronized as $s
\bar{s} ( \bar{u} u + \bar{d} d + \bar{s} s ) \equiv ( \phi \cdot \phi
)_{3 3}$, where
\begin{align}
( \phi \cdot \phi )_{3 3}
& =
K^{-} K^{+} + \bar{K}^{0} K^{0} 
+ \frac{1}{3} \eta \eta .
\label{eq:phiphi33}
\end{align}
Here and in the following we omit the $\eta ^{\prime}$ contribution
since $\eta ^{\prime}$ is irrelevant to the description of the scalar
mesons due to its large mass.  In similar manners, the $d \bar{s}$, $s
\bar{d}$, $d \bar{d}$, $s \bar{u}$, and $d \bar{u}$ pairs get
hadronized as
\begin{align}
( \phi \cdot \phi )_{2 3} 
= \pi ^{-} K^{+} - \frac{1}{\sqrt{2}} \pi ^{0} K^{0} ,
\end{align}
\begin{align}
( \phi \cdot \phi )_{3 2}
= & K^{-} \pi ^{+} - \frac{1}{\sqrt{2}} \bar{K}^{0} \pi ^{0} ,
\end{align}
\begin{align}
( \phi \cdot \phi )_{2 2}
& = \pi ^{-} \pi ^{+} + \frac{1}{2} \pi ^{0} \pi ^{0} 
+ \frac{1}{3} \eta \eta 
- \sqrt{\frac{2}{3}} \pi ^{0} \eta 
+ K^{0} \bar{K}^{0} ,
\end{align}
\begin{align}
( \phi \cdot \phi )_{3 1} 
= \frac{1}{\sqrt{2}} \pi ^{0} K^{-} + \pi ^{-} \bar{K}^{0} ,
\end{align}
and
\begin{align}
( \phi \cdot \phi )_{2 1} 
= \frac{2}{\sqrt{3}} \pi ^{-} \eta + K^{0} K^{-} ,
\end{align}
respectively.

By using these weights, we can express the hadronization amplitude for
the scalar mesons, $V_{\rm had}^{(s)}$, in terms of two pseudoscalar
mesons.  For instance, we want to reconstruct $f_{0} (500)$ and $f_{0}
(980)$ from the $\pi ^{+} \pi ^{-}$ system in the $D_{s}^{+} \to \pi
^{+} \pi ^{-} \, l^{+} \, \nu _{l}$ decay.  Because of the quark
configuration in the parent particle $D_{s}^{+}$, in this decay the
$\pi ^{+} \pi ^{-}$ system should be obtained from the hadronization
of the $s \bar{s}$ pair and the rescattering process for two
pseudoscalar mesons, as seen in Fig.~\ref{fig:3}, with the weight in
Eq.~\eqref{eq:phiphi33}.  Therefore, for the $D_{s}^{+} \to \pi ^{+}
\pi ^{-} \, l^{+} \, \nu _{l}$ decay mode we can express the
hadronization amplitude with a prefactor $C$ and the
Cabibbo--Kobayashi--Maskawa matrix elements $V_{c s}$ as
\begin{align}
& V_{\rm had}^{(s)} [ D_{s}^{+}, \, \pi ^{+} \pi ^{-}]
= C V_{c s} \left ( G_{K^{+} K^{-}} T_{K^{+} K^{-} \to \pi ^{+} \pi ^{-}} 
\phantom{\frac{1}{2}}
\right .
\notag \\
& \left . + G_{K^{0} \bar{K}^{0}} T_{K^{0} \bar{K}^{0} \to \pi ^{+} \pi ^{-}} 
+ \frac{1}{3} \cdot 2 \cdot \frac{1}{2} 
G_{\eta \eta} T_{\eta \eta \to \pi ^{+} \pi ^{-}} 
\right ) .
\end{align}
In this equation, the decay mode is abbreviated as $[ D_{s}^{+}, \,
\pi ^{+} \pi ^{-}]$, and $G$ and $T$ are the loop function and
scattering amplitude of two pseudoscalar mesons, respectively, whose
formulation are given in Sec.~\ref{sec:ChUA}.  We have introduced
extra factors $2$ and $1/2$ for the identical particles $\eta \eta$.
The former factor $2$ comes from the two ways of annihilating the
$\eta \eta$ operator in Eq.~\eqref{eq:phiphi33} by the $| \eta \eta
\rangle$ state as in the usual manner for effective Lagrangians, while
the latter one $1/2$ is the symmetry factor for the $\eta \eta$ loop.
The scalar mesons $f_{0} (500)$ and $f_{0} (980)$ appear in the
rescattering process and exist in the scattering amplitude $T$ for two
pseudoscalar mesons.  It is important that this is a Cabibbo favored
process with $V_{c s}$.  Furthermore, since the $s \bar{s}$ pair is
hadronized, this is sensitive to the component of the strange quark in
the scalar mesons.  In this study we assume that $C$ is a constant,
and hence the hadronization amplitude $V_{\rm had}^{(s)}$ is a
function only of the invariant mass of two pseudoscalar mesons.  Here
we emphasize that the prefactor $C$ should be common to all reactions
for scalar meson productions, because in the hadronization the SU(3)
flavor symmetry is reasonable, i.e., the light quark--antiquark pair
$q_{f} \bar{q}_{f^{\prime}}$ hadronizes in the same way regardless of
the quark flavor $f$.  In this sense we obtain
\begin{align}
& V_{\rm had}^{(s)} [ D_{s}^{+}, \, K^{+} K^{-}]
= C V_{c s} \left ( 1 + G_{K^{+} K^{-}} T_{K^{+} K^{-} \to K^{+} K^{-}} 
\phantom{\frac{1}{2}}
\right .
\notag \\
& \left . + G_{K^{0} \bar{K}^{0}} T_{K^{0} \bar{K}^{0} \to K^{+} K^{-}} 
+ \frac{1}{3} \cdot 2 \cdot \frac{1}{2} 
G_{\eta \eta} T_{\eta \eta \to K^{+} K^{-}} 
\right ) ,
\end{align}
for the $D_{s}^{+} \to K^{+} K^{-} \, l^{+} \, \nu _{l}$ decay.  In
this case we have to take into account the direct production of the
two pseudoscalar mesons without rescattering (the first diagram in
Fig.~\ref{fig:3}), which results in the unity in the parentheses.  On
the other hand, for the $D_{s}^{+} \to \pi ^{-} K^{+} \, l^{+} \, \nu
_{l}$ decay mode the $\pi ^{-} K^{+}$ system should be obtained from
the hadronization of $d \bar{s}$ and hence this is a Cabibbo
suppressed decay mode.  The hadronization amplitude is expressed as
\begin{align}
& V_{\rm had}^{(s)} [ D_{s}^{+}, \, \pi ^{-} K^{+}]
= C V_{c d} \left ( 1 + G_{\pi ^{-} K^{+}} T_{\pi ^{-} K^{+} \to \pi ^{-} K^{+}}
\phantom{\frac{1}{\sqrt{2}}}
\right .
\notag \\
& \left . 
\quad \quad \quad \quad 
- \frac{1}{\sqrt{2}} G_{\pi ^{0} K^{0}} T_{\pi ^{0} K^{0} \to \pi ^{-} K^{+}}
\right ) .
\label{eq:Ds_pipi}
\end{align}
In a similar manner we can construct every hadronization amplitude for
the scalar meson.  The resulting expressions are as follows:
\begin{align}
& V_{\rm had}^{(s)} [ D^{+}, \, \pi ^{+} \pi ^{-}]
= C V_{c d} \left ( 1 + 
G_{\pi ^{+} \pi ^{-}} T_{\pi ^{+} \pi ^{-} \to \pi ^{+} \pi ^{-}} 
\phantom{\frac{1}{2}}
\right .
\notag \\ 
& 
+ \frac{1}{2} \cdot 2 \cdot \frac{1}{2} 
G_{\pi ^{0} \pi ^{0}} T_{\pi ^{0} \pi ^{0} \to \pi ^{+} \pi ^{-}} 
+ \frac{1}{3} \cdot 2 \cdot \frac{1}{2} 
G_{\eta \eta} T_{\eta \eta \to \pi ^{+} \pi ^{-}} 
\notag \\ 
& 
\left . \! \! \! \! \! \!
\phantom{\frac{1}{2}}
+ G_{K^{0} \bar{K}^{0}} T_{K^{0} \bar{K}^{0} \to \pi ^{+} \pi ^{-}} 
\right ) ,
\end{align}
\begin{align}
& V_{\rm had}^{(s)} [ D^{+}, \, \pi ^{0} \eta ]
= C V_{c d} \left ( - \sqrt{\frac{2}{3}} 
- \sqrt{\frac{2}{3}} 
G_{\pi ^{0} \eta} T_{\pi ^{0} \eta \to \pi ^{0} \eta} 
\right .
\notag \\
& \left . \phantom{\sqrt{\frac{2}{3}}} 
  \quad \quad \quad \quad 
  + G_{K^{0} \bar{K}^{0}} T_{K^{0} \bar{K}^{0} \to \pi ^{0} \eta}
\right ) ,
\end{align}
\begin{align}
& V_{\rm had}^{(s)} [ D^{+}, \, K^{+} K^{-}]
= C V_{c d} \left ( G_{\pi ^{+} \pi ^{-}} T_{\pi ^{+} \pi ^{-} \to K^{+} K^{-}} 
\phantom{\sqrt{\frac{1}{2}}}
\right .
\notag \\ 
& 
+ \frac{1}{2} \cdot 2 \cdot \frac{1}{2} 
G_{\pi ^{0} \pi ^{0}} T_{\pi ^{0} \pi ^{0} \to K^{+} K^{-}} 
+ \frac{1}{3} \cdot 2 \cdot \frac{1}{2} 
G_{\eta \eta} T_{\eta \eta \to K^{+} K^{-}} 
\notag \\ 
& 
\left . \!
- \sqrt{\frac{2}{3}} 
G_{\pi ^{0} \eta} T_{\pi ^{0} \eta \to K^{+} K^{-}} 
+ G_{K^{0} \bar{K}^{0}} T_{K^{0} \bar{K}^{0} \to K^{+} K^{-}} 
\right ) ,
\end{align}
\begin{align}
& V_{\rm had}^{(s)} [ D^{+}, \, \pi ^{+} K^{-}]
= C V_{c s} \left ( 1 + G_{\pi ^{+} K^{-}} T_{\pi ^{+} K^{-} \to \pi ^{+} K^{-}}
\phantom{\frac{1}{\sqrt{2}}}
\right .
\notag \\ 
& 
\left . 
  \quad \quad \quad \quad 
- \frac{1}{\sqrt{2}} 
G_{\pi ^{0} \bar{K}^{0}} T_{\pi ^{0} \bar{K}^{0} \to \pi ^{+} K^{-}} 
\right ) ,
\end{align}
\begin{align}
& V_{\rm had}^{(s)} [ D^{0}, \, \pi ^{-} \eta ]
= C V_{c d} \left ( \frac{2}{\sqrt{3}} 
+ \frac{2}{\sqrt{3}} 
G_{\pi ^{-} \eta} T_{\pi ^{-} \eta \to \pi ^{-} \eta} 
\right .
\notag \\
& \left . \phantom{\frac{2}{\sqrt{3}}} 
  \quad \quad \quad \quad 
  + G_{K^{0} K^{-}} T_{K^{0} K^{-} \to \pi ^{-} \eta}
\right ) ,
\end{align}
\begin{align}
& V_{\rm had}^{(s)} [ D^{0}, \, K^{0} K^{-} ]
= C V_{c d} \left ( 1 
+ \frac{2}{\sqrt{3}} 
G_{\pi ^{-} \eta} T_{\pi ^{-} \eta \to K^{0} K^{-}} 
\right .
\notag \\
& \left . \phantom{\frac{2}{\sqrt{3}}} 
  \quad \quad \quad \quad 
  + G_{K^{0} K^{-}} T_{K^{0} K^{-} \to K^{0} K^{-}}
\right ) ,
\end{align}
\begin{align}
& V_{\rm had}^{(s)} [ D^{0}, \, \pi ^{-} \bar{K}^{0} ]
= C V_{c s} \left ( 1 
+ \frac{1}{\sqrt{2}} 
G_{\pi ^{0} K^{-}} T_{\pi ^{0} K^{-} \to \pi ^{-} \bar{K}^{0}} 
\right .
\notag \\
& \left . \phantom{\frac{1}{\sqrt{2}}} 
  \quad \quad \quad \quad 
  + G_{\pi ^{-} \bar{K}^{0}} T_{\pi ^{-} \bar{K}^{0} \to \pi ^{-} \bar{K}^{0}}
\right ) .
\end{align}
The hadronization amplitudes $V_{\rm had}^{(s)} [ D_{s}^{+}, \, \pi
^{-} K^{+}]$, $V_{\rm had}^{(s)} [ D^{+}, \, \pi ^{+} K^{-}]$, and
$V_{\rm had}^{(s)} [ D^{0}, \, \pi ^{-} \bar{K}^{0} ]$ are further
simplified by using the isospin symmetry as
\begin{equation}
V_{\rm had}^{(s)} [ D_{s}^{+}, \, \pi
^{-} K^{+}] = C V_{c d} A_{\pi K} ,
\end{equation}
\begin{equation}
V_{\rm had}^{(s)} [ D^{+}, \, \pi ^{+} K^{-}] = 
V_{\rm had}^{(s)} [ D^{0}, \, \pi ^{-} \bar{K}^{0}] =
C V_{c s} A_{\pi K} ,
\end{equation}
where $A_{\pi K}$ is a function of the invariant mass of two
pseudoscalar mesons and is defined with the scattering amplitude in the
isospin basis as
\begin{equation}
A_{\pi K} \equiv 1 + G_{\pi K} T_{\pi K (I = 1/2) \to \pi K (I = 1/2)} .
\end{equation}
In a similar manner, we simplify the hadronization amplitudes $V_{\rm
  had}^{(s)} [ D^{+}, \, \pi ^{0} \eta ]$, $V_{\rm had}^{(s)} [ D^{0},
\, \pi ^{-} \eta ]$, and $V_{\rm had}^{(s)} [ D^{0}, \, K^{0} K^{-} ]$
as
\begin{equation}
V_{\rm had}^{(s)} [ D^{+}, \, \pi ^{0} \eta ] 
= - \frac{1}{\sqrt{2}} V_{\rm had}^{(s)} [ D^{0}, \, \pi ^{-} \eta ] 
= - \sqrt{\frac{2}{3}} C V_{c d} B_{\pi \eta} ,
\end{equation}
\begin{equation}
V_{\rm had}^{(s)} [ D^{0}, \, K^{0} K^{-} ]
= C V_{c d} B_{K \bar{K}} ,
\end{equation}
with 
\begin{equation}
B_{\pi \eta} \equiv 1 + G_{\pi \eta} T_{\pi \eta \to \pi \eta} 
- \frac{\sqrt{3}}{2} G_{K \bar{K}} T_{K \bar{K} (I = 1) \to \pi \eta} .
\end{equation}
\begin{align}
B_{K \bar{K}} \equiv & 1 + G_{K \bar{K}} T_{K \bar{K} (I = 1) \to K \bar{K} (I = 1)} 
\notag \\
& - \frac{2}{\sqrt{3}} G_{\pi \eta} T_{\pi \eta \to K \bar{K} (I = 1)} .
\end{align}

From the above expressions one can easily specify Cabibbo favored and
suppressed processes for the semileptonic decays into two pseudoscalar
mesons, which are listed in Table~\ref{tab:mode}.

Finally we note that the use of a constant $C$ factor in our approach
gets support from the work of Ref.~\cite{Kang:2013jaa}.  The
evaluation of the matrix elements in these processes is difficult and
problematic.  There are however some cases where the calculations can
be kept under control.  For the case of small recoil, namely when
final pseudoscalars move slow, it can be explored in the heavy meson
chiral perturbation theory~\cite{Manohar:2000dt}.  Detailed
calculations for the case of semileptonic decay are done
in~\cite{Kang:2013jaa}.  There one can see that for large values of
the invariant mass of the lepton system the form factors can be
calculated and the relevant ones in $s$ wave that we need here are
smooth in the range of the invariant masses of the pairs of mesons
that we use here.  To be able to use this behaviour we should prove
that in our case the invariant masses of the lepton pair are large,
but indeed, it was shown in the study of the semileptonic $B$
decays~\cite{Navarra:2015iea} (and can be done also here) that the
mass distribution of the lepton pair accumulates at the upper end of
the phase space.  There is also another limit, at large recoil, where
an approach that combines both hard-scattering and low-energy
interactions has been developed is also
available~\cite{Meissner:2013hya}, but this is not the case here.

\subsubsection{Vector mesons}

Next we consider processes with the vector mesons in the final state.
As we have already mentioned, hadronization with an extra $\bar{q} q$
is unnecessary for the vector mesons.  As a consequence, we can
formulate the hadronization amplitude for vector mesons, $V_{\rm
  had}^{(v)}$, in a very simple way.

In order to see this, we consider the semileptonic decay $D_{s}^{+}
\to \phi (1020) \, l^{+} \, \nu _{l}$ as an example.  The decay
process is diagrammatically represented in Fig.~\ref{fig:12}(a), and the
hadronization amplitude $V_{\rm had}^{(v)}$ can be expressed with a
prefactor $C^{\prime}$ and the Cabibbo--Kobayashi--Maskawa matrix
element $V_{c s}$ as
\begin{equation}
V_{\rm had}^{(v)} [ D_{s}^{+}, \, \phi ] = C^{\prime} V_{c s} ,
\end{equation}
where the decay mode is abbreviated as $[ D_{s}^{+}, \, \phi ]$ in the
equation.  Here we emphasize that the prefactor $C^{\prime}$ should be
common to all reactions for vector meson productions, as in the case
of the scalar meson productions, because the SU(3) flavor symmetry is
reasonable in the hadronization, i.e., the light quark--antiquark pair
$q_{f} \bar{q}_{f^{\prime}}$ hadronizes in the same way regardless of
the quark flavor $f$.  We further assume that $C^{\prime}$ is a
constant again.  This formulation is straightforwardly applied to
other vector meson productions and we obtain the hadronization
amplitude for vector mesons:
\begin{equation}
V_{\rm had}^{(v)} [ D_{s}^{+}, \, K^{\ast 0} ] = C^{\prime} V_{c d} ,
\end{equation}
\begin{equation}
V_{\rm had}^{(v)} [ D^{+}, \, \bar{K}^{\ast 0} ] = C^{\prime} V_{c s} ,
\end{equation}
\begin{equation}
V_{\rm had}^{(v)} [ D^{+}, \, \rho ^{0} ] 
= - \frac{1}{\sqrt{2}} C^{\prime} V_{c d} ,
\end{equation}
\begin{equation}
V_{\rm had}^{(v)} [ D^{+}, \, \omega ] 
= \frac{1}{\sqrt{2}} C^{\prime} V_{c d} ,
\end{equation}
\begin{equation}
V_{\rm had}^{(v)} [ D^{0}, \, K^{\ast -} ] = - C^{\prime} V_{c s} ,
\end{equation}
\begin{equation}
V_{\rm had}^{(v)} [ D^{0}, \, \rho ^{-} ] 
= C^{\prime} V_{c d} ,
\end{equation}
where we have used $K^{\ast}$, $\rho$ and $\omega$ states in the
isospin basis summarized in Appendix~\ref{app:1}.  We note that these
equations clearly indicate Cabibbo favored and suppressed processes
with the Cabibbo--Kobayashi--Maskawa matrix elements $V_{c s}$ and $V_{c
  d}$, respectively.

\subsection{Scattering amplitudes of two pseudoscalar mesons in chiral
  unitary approach}
\label{sec:ChUA}

For the scattering amplitude of two pseudoscalar mesons, we employ the
so-called chiral unitary approach~\cite{Oller:1997ti, Kaiser:1998fi,
  Locher:1997gr, Oller:1997ng, Oller:1998hw, Oller:1998zr,
  Nieves:1999bx, Pelaez:2006nj, Albaladejo:2008qa}, which we briefly
explain in this subsection.  In this approach we solve a
coupled-channels Bethe--Salpeter equation in an algebraic form
\begin{equation}
T_{i j} (s) = V_{i j} (s) + \sum _{k} V_{i k} (s) G_{k} (s) T_{k j} (s) ,
\label{eq:BSEq}
\end{equation}
where $i$, $j$, and $k$ are channel indices, $s$ is the Mandelstam
variable of the scattering, $V$ is the interaction kernel, and $G$ is
the two-body loop function.  For the hadronization in the previous
subsection we need three types of coupled-channels systems: the $(Q,
\, S) = (0, \, 0)$ system, for which we introduce six channels labeled
by the indices $i = 1$, $\ldots$, $6$ in the order $\pi ^{+} \pi
^{-}$, $\pi ^{0} \pi ^{0}$, $K^{+} K^{-}$, $K^{0} \bar{K}^{0}$, $\eta
\eta$, and $\pi ^{0} \eta$, the $K \bar{K} (I=1)$-$\pi \eta$ system,
and the $\pi K (I=1/2)$-$\eta K$ system.

In this study the interaction kernel $V_{i j} = V_{j i}$ is taken as
the simplest one, that is, the leading-order $s$-wave interaction
obtained from the chiral perturbation theory.  The interaction kernel
for $(Q, \, S) = (0, \, 0)$ is summarized as
\begin{equation}
\begin{split}
& V_{11} = 2 V_{13} = 2 V_{14} = 2 \sqrt{2} V_{23} = 2 \sqrt{2} V_{24}
\\
& = V_{33} = 2 V_{34} = V_{44} 
= - \frac{s}{2 f^{2}} , 
\\
& V_{12} = - \frac{s - m_{\pi}^{2}}{\sqrt{2} f^{2}} , 
\\
& V_{15} = \frac{\sqrt{2}}{3} V_{22} = \sqrt{2} V_{25} 
= \frac{1}{\sqrt{2}} V_{66}
= - \frac{m_{\pi}^{2}}{3 \sqrt{2} f^{2}} , 
\\
& V_{16} = V_{26} = V_{56} = 0 , 
\\
& V_{35} = V_{45} 
= - \frac{9 s - 2 m_{\pi}^{2} - 6 m_{\eta}^{2}}{12 \sqrt{2} f^{2}} ,
\\
& V_{36} = - V_{46} 
= - \frac{9 s - m_{\pi}^{2} - 8 m_{K}^{2} - 3 m_{\eta}^{2}}{12 \sqrt{3} f^{2}} ,
\\
& V_{55} = \frac{7 m_{\pi}^{2} - 16 m_{K}^{2}}{18 f^{2}} ,
\end{split}
\label{eq:Vint} 
\end{equation}
where $f$ is the pion decay constant.  One must remember that in the
chiral unitary approach when calculating $T = (1 - V G)^{-1} V$ one
uses the unitary normalization $(1/\sqrt{2}) | \pi ^{0} \pi ^{0}
\rangle$ and $(1/\sqrt{2}) | \eta \eta \rangle$ for identical
particles, which allows to use the general formula in coupled
channels.  At the end the good normalization of the external particles
must be restored and these are the amplitudes that appear to
Eq.~\eqref{eq:Ds_pipi} and following ones.

For the $K \bar{K} (I = 1)$-$\pi \eta$ scattering, the interaction
kernel can be written in terms of the interaction kernel for the $(Q,
\, S) = (0, \, 0)$ system shown above (see the isospin basis
summarized in Appendix~\ref{app:1}):
\begin{equation}
V_{K \bar{K} (I = 1) \to K \bar{K} (I = 1)}
= \frac{1}{2} \left ( V_{3 3} - 2 V_{3 4} + V_{4 4} \right ) , 
\end{equation}
\begin{equation}
V_{K \bar{K} (I = 1) \leftrightarrow \pi \eta}
= - \frac{1}{\sqrt{2}} \left ( V_{3 6} - V_{4 6} \right ) , 
\end{equation}
\begin{equation}
V_{\pi \eta \to \pi \eta}
= V_{6 6} . 
\end{equation}
For the $\pi K (I = 1/2)$-$\eta K$ scattering, the interaction kernel
is expressed as
\begin{align}
V_{\pi K (I = 1/2) \to \pi K (I = 1/2)} = & \frac{1}{8 s f^{2}}  
\left [ - 5 s^{2} + 2 ( m_{\pi}^{2} + m_{K}^{2} ) s \right .
\notag \\
& \left . + 3 ( m_{\pi} - m_{K} )^{2} \right ] , 
\end{align}
\begin{align}
V_{\pi K (I = 1/2) \leftrightarrow \eta K} = & \frac{1}{24 s f^{2}} \left [ 
9 s^{2} - ( 7 m_{\pi}^{2} + 2 m_{K}^{2} + 3 m_{\eta}^{2} ) s 
\right . 
\notag \\
& \left . 
  - 9 ( m_{\pi}^{2} - m_{K}^{2} ) ( m_{K}^{2} - m_{\eta}^{2} )  \right ] , 
\end{align}
\begin{align}
V_{\eta K \to \eta K} = & \frac{1}{24 s f^{2}} \left [ 
  9 s^{2} + 2 ( 2 m_{\pi}^{2} - 9 m_{K}^{2} - 3 m_{\eta}^{2} ) s 
\right .
\notag \\
& \left .  + 9 ( m_{K} - m_{\eta} )^{2} 
\right ] . 
\end{align}

For the loop function $G$, on the other hand, we use the following
expression:
\begin{equation}
G_{i} ( s ) \equiv 
i \int \frac{d ^{4} q}{(2 \pi)^{4}} 
\frac{1}{q^{2} - m_{i}^{2} + i 0} 
\frac{1}{(P - q)^{2} - m_{i}^{\prime 2} + i 0} .
\end{equation}
where $P^{\mu} = (\sqrt{s}, \, \bm{0})$ and $m_{i}$ and
$m_{i}^{\prime}$ are masses of pseudoscalar mesons in channel $i$.  In
this study we employ a three-dimensional cut-off $q_{\rm max}$ as
\begin{equation}
G_{i} (s) = \int \frac{d^{3} q}{(2 \pi )^{3}}
\frac{\omega _{i}(\bm{q}) + \omega _{i}^{\prime}(\bm{q})}
{2 \omega _{i}(\bm{q}) \omega _{i}^{\prime}(\bm{q})}
\frac{\theta ( q_{\rm max} - |\bm{q}| )}
{s - [\omega _{i}(\bm{q}) + \omega _{i}^{\prime}(\bm{q})]^{2}
+ i 0} ,
\label{eq:Gloop}
\end{equation}
In this expression we have performed the $q^{0}$ integral and $\omega
_{i}(\bm{q}) \equiv \sqrt{m_{i}^{2} + \bm{q}^{2}}$ and $\omega
_{i}^{\prime}(\bm{q}) \equiv \sqrt{m_{i}^{\prime 2} + \bm{q}^{2}}$ are
the on-shell energies.

In this framework, with a small number of free parameters we can
reproduce experimental observables of meson--meson scatterings fairly
well.  In this study we take the model parameters of the chiral
unitary approach as $f = 93 \mev$ and $q_{\rm max} = 600 \mev$, which
dynamically generates resonance poles in the complex energy plane:
$453 - 253 i \mev$ for $f_{0} (500)$, $982 - 5 i \mev$ for $f_{0}
(980)$, and $721 - 236 i \mev$ for $K_{0}^{\ast} (800)$.  The $a_{0}
(980)$ appears as a cusp at the $K \bar{K}$ threshold.

\section{Numerical results}
\label{sec:results}

Now let us calculate the semileptonic decay widths of $D$ mesons into
scalar and vector mesons.  As we have formulated, we have only one
model parameter for scalar and vector meson productions, respectively.
Namely one can calculate the decay widths of the scalar meson
productions with one common parameter $C$, and similarly $C^{\prime}$
for the vector meson productions.

First we consider the scalar meson production in
Sec.~\ref{sec:scalar}, and then move to the vector meson production in
Sec.~\ref{sec:vector}.  Finally in Sec.~\ref{sec:compare} we compare
the two contributions of the mass distributions from the scalar and
vector mesons.

\subsection{Production of scalar mesons}
\label{sec:scalar}

In order to calculate the branching fractions of the scalar meson
productions, we first fix the prefactor constant $C$ so as to
reproduce the experimental branching fraction which has the smallest
experimental error for the process with the $s$-wave two pseudoscalar
mesons, that is, $\mathcal{B}[D^{+} \to ( \pi ^{+} K^{-}
)_{s\text{-wave}} e^{+} \nu _{e}] = (2.32 \pm 0.10) \times 10^{-3}$.
By integrating the differential decay width, or mass distribution, $d
\Gamma _{4} / d M_{\rm inv}^{(hh)}$ in an appropriate range, in the
case of $\pi ^{+} K^{-}$ [$m_{\pi} + m_{K}$, $1 \gev$], we find that
$C = 4.597$ can reproduce the branching fraction of $( \pi ^{+} K^{-}
)_{s\text{-wave}} e^{+} \nu _{e}$.

\begin{figure}[!t]
  \centering
  \Psfig{8.6cm}{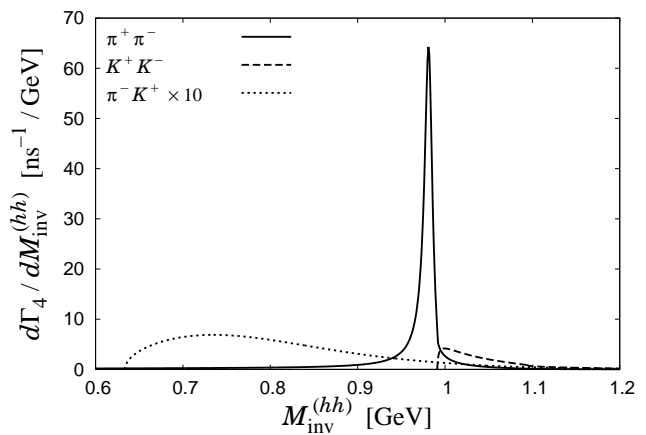}
  \caption{Meson--meson invariant mass distributions for the
    semileptonic decay $D_{s}^{+} \to P P e^{+} \nu _{e}$ with $P P =
    \pi ^{+} \pi ^{-}$, $K^{+} K^{-}$, and $\pi ^{-} K^{+}$ in $s$
    wave.  We multiply the $\pi ^{-} K^{+}$ mass distribution, which
    is a Cabibbo suppressed process, by $10$. }
\label{fig:dG4_Ds}
\end{figure}

\begin{figure}[!t]
  \centering
  \Psfig{8.6cm}{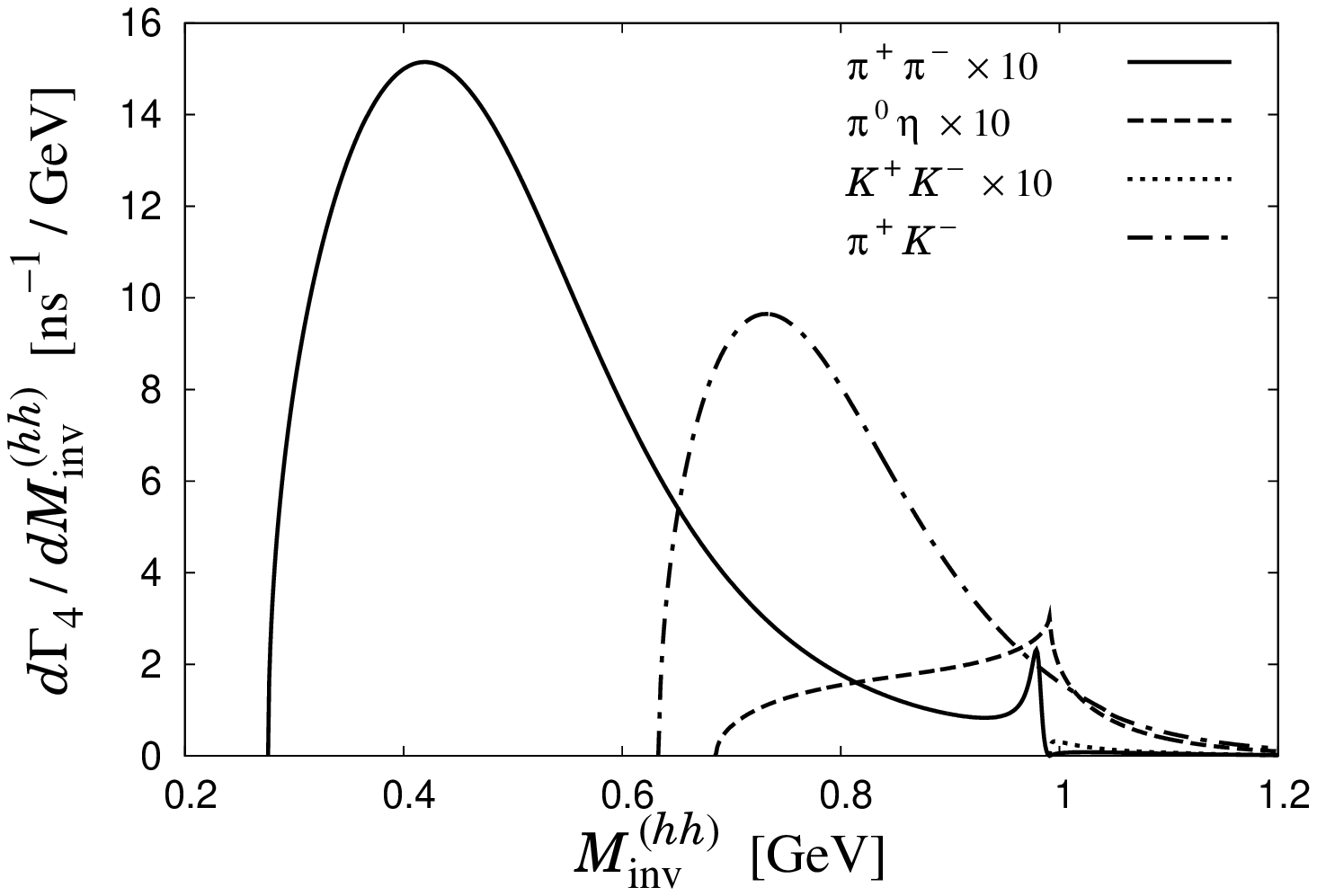}
  \caption{Meson--meson invariant mass distributions for the
    semileptonic decay $D^{+} \to P P e^{+} \nu _{e}$ with $P P = \pi
    ^{+} \pi ^{-}$, $\pi ^{0} \eta$, $K^{+} K^{-}$, and $\pi ^{+}
    K^{-}$ in $s$ wave.  We multiply the $\pi ^{+} \pi ^{-}$, $\pi
    ^{0} \eta$, and $K^{+} K^{-}$ mass distributions, which are
    Cabibbo suppressed processes, by $10$. }
\label{fig:dG4_Dp}
\end{figure}

\begin{figure}[!t]
  \centering
  \Psfig{8.6cm}{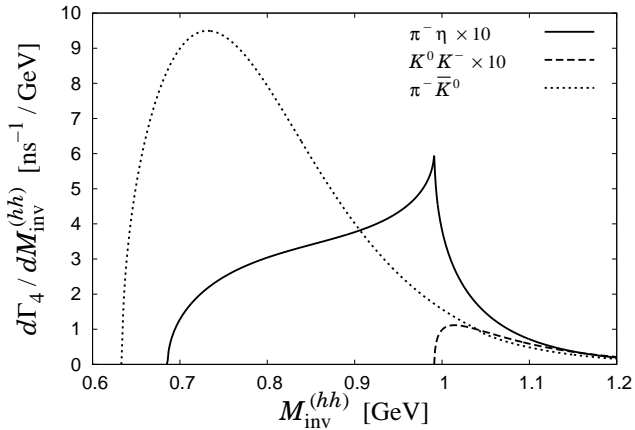}
  \caption{Meson--meson invariant mass distributions for the
    semileptonic decay $D^{0} \to P P e^{+} \nu _{e}$ with $P P = \pi
    ^{-} \eta$, $K^{0} K^{-}$, and $\pi ^{-} \bar{K}^{0}$ in $s$ wave.
    We multiply the $\pi ^{-} \eta$ and $K^{0} K^{-}$ mass
    distributions, which are Cabibbo suppressed processes, by $10$. }
\label{fig:dG4_Dz}
\end{figure}

By using the common prefactor $C = 4.597$, we can calculate the mass
distributions of two pseudoscalar mesons in $s$ wave for all scalar
meson modes, which are plotted in Figs.~\ref{fig:dG4_Ds},
\ref{fig:dG4_Dp}, and \ref{fig:dG4_Dz} for $D_{s}^{+}$, $D^{+}$, and
$D^{0}$ semileptonic decays, respectively.  We show the mass
distributions with the lepton flavor $l = e$; the contribution from $l
= \mu$ is almost the same as that from $l = e$ in each meson--meson
mode due to the small lepton masses.  In each figure we multiply the
mass distributions which are Cabibbo suppressed processes by $10$ so
that we can easily compare the shape of the mass distributions.  As
one can see, the largest value of the mass distribution $d \Gamma _{4}
/ d M_{\rm inv}^{(hh)}$ is obtained in the $D_{s}^{+} \to \pi ^{+} \pi
^{-} e^{+} \nu _{e}$ process, in which we can see a clear $f_{0}
(980)$ peak.  It is interesting to note that in the $D_{s}^{+} \to \pi
^{+} \pi ^{-} e^{+} \nu _{e}$ process we find a clear $f_{0} (980)$
signal while the $f_{0} (500)$ contribution is negligible, which
strongly indicates a substantial fraction of the strange quarks in the
$f_{0} (980)$ meson, as we will discuss later.  For the $D_{s}^{+}$
semileptonic decay we also observe a rapid enhancement of the $K^{+}
K^{-}$ mass distribution at the threshold, as a tail of the $f_{0}
(980)$ contribution, although its height is much smaller than the $\pi
^{+} \pi ^{-}$ peak.  For the $D^{+}$ and $D^{0}$ semileptonic decays,
we can see the $\pi ^{+} K^{-}$ and $\pi ^{-} \bar{K}^{0}$ as Cabibbo
favored processes, respectively.  We note that the $\pi ^{+} K^{-}$
and $\pi ^{-} \bar{K}^{0}$ mass distributions are almost the same due
to isospin symmetry.  It is interesting to see that the shape of the
$\pi ^{+} K^{-}$ and $\pi ^{-} \bar{K}^{0}$ mass distributions is
determined by, in addition to the $K_{0}^{\ast} (800)$ resonance, the
kinetic factor of the squared decay amplitude.  Namely, we have the
matrix element of Eq.~\eqref{eq:amp2} that is roughly proportional to
$| \bm{p}_{\nu}|^{2}$ and this momentum gets bigger the smaller the
meson--meson invariant mass.  This kinetic factor of the squared decay
amplitude affects the $\pi ^{+} \pi ^{-}$ distribution in the $D^{+}$
semileptonic decay in a similar manner, and also provides more weight
at low invariant masses for the shape for $\pi \eta$ in
Figs.~\ref{fig:dG4_Dp} and \ref{fig:dG4_Dz} than the $\pi ^{0} \eta$
distributions in the $D^{0} \to \bar{K}^{0} \pi ^{0} \eta$ decay
evaluated in Ref.~\cite{Xie:2014tma}.  The $\pi \eta$ mass
distributions in Figs.~\ref{fig:dG4_Dp} and \ref{fig:dG4_Dz} of the
$D^{+}$ and $D^{0}$ decays show peaks corresponding to $a_{0} (980)$,
but its peak is not high compared to the $f_{0} (980)$ peak in the
$\pi ^{+} \pi ^{-}$ mass distribution of the $D_{s}^{+}$ decay since
they are obtained in Cabibbo suppressed processes.  The $D^{+} \to \pi
^{+} \pi ^{-} e^{+} \nu _{e}$ decay is Cabibbo suppressed and it has a
large contribution from the $f_{0} (500)$ formation and a small one of
the $f_{0} (980)$, similar to what is found in the $\bar{B}^{0} \to J
/ \psi \pi ^{+} \pi ^{-}$ decay in Ref.~\cite{Liang:2014tia}.  A
different way to account for the $PP$ distribution is by means of
dispersion relations, as used in Ref.~\cite{Kang:2013jaa} in the
semileptonic decay of $B$, where the $\pi ^{+} \pi ^{-}$ $s$-wave
distribution has a shape similar to ours.

\begin{figure}[!t]
  \centering
  \Psfig{8.6cm}{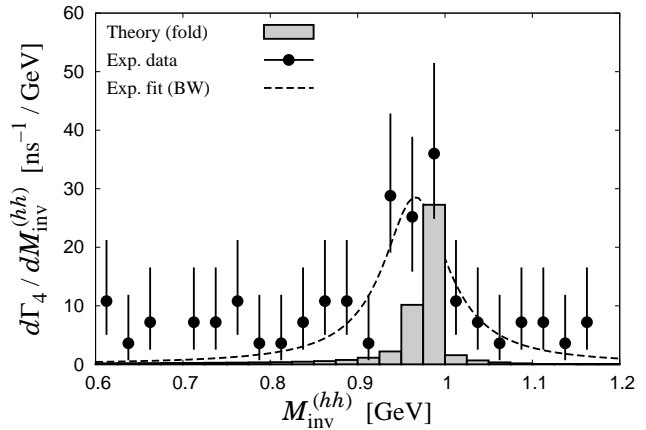}
  \caption{$\pi ^{+} \pi ^{-}$ invariant mass distribution for the
    semileptonic decay $D_{s}^{+} \to \pi ^{+} \pi ^{-} e^{+} \nu
    _{e}$.  The theoretical calculation is folded with the size of
    experimental bins, $25 \mev$.  The experimental data are taken from
    Ref.~\cite{Ecklund:2009aa} and are scaled so that the fitted
    Breit--Wigner distribution (dashed line) reproduces the branching
    fraction of $\mathcal{B}[D_{s}^{+} \to f_{0} (980) e^{+} \nu _{e},
    \, f_{0} (980) \to \pi ^{+} \pi ^{-}] = 0.2 \%$ by the Particle
    Data Group (see Table~\ref{tab:Br}). }
\label{fig:hist}
\end{figure}

The theoretical $\pi ^{+} \pi ^{-}$ mass distribution of the
semileptonic decay $D_{s} \to \pi ^{+} \pi ^{-} e^{+} \nu _{e}$ is
compared with the experimental data~\cite{Ecklund:2009aa} in
Fig.~\ref{fig:hist}.  We note that we plot the figure in unit of
$\text{ns}^{-1} / \text{GeV}$, not in arbitrary units.  The
theoretical mass distribution is folded with $25 \mev$ spans since the
experimental data are collected in bins of $25 \mev$.  The
experimental data, on the other hand, are scaled so that the fitted
Breit--Wigner distribution reproduces the branching fraction of
$\mathcal{B}[D_{s}^{+} \to f_{0} (980) e^{+} \nu _{e}, \, f_{0} (980)
\to \pi ^{+} \pi ^{-}] = 0.2 \%$~\cite{Agashe:2014kda}.  The mass and
width of the Breit--Wigner distribution are fixed as $M = 966 \mev$
and $\Gamma = 89 \mev$, respectively, taken from
Ref.~\cite{Ecklund:2009aa}.  In Fig.~\ref{fig:hist} we can see a
qualitative correspondence between the theoretical and experimental
signals of $f_{0} (980)$.  We emphasize that, both in experimental and
theoretical results, the $\pi ^{+} \pi ^{-}$ mass distribution shows a
clear $f_{0} (980)$ signal while the $f_{0} (500)$ contribution is
negligible.  This strongly indicates that the $f_{0} (980)$ has a
substantial fraction of the strange quarks while the $f_{0} (500)$ has
a negligible strange quark component.  It is interesting to recall
that the appearance of the $f_{0} (980)$ in the case one has a
hadronized $s \bar{s}$ component at the end, and no signal of the
$f_{0} (500)$, is also observed in $B_{s}^{0}$ and $B^{0}$ decays in
Refs.~\cite{Aaij:2011fx, Li:2011pg, Aaltonen:2011nk, Abazov:2011hv,
  LHCb:2012ae}.  The explanation of this feature along the lines used
in the present work was given in Ref.~\cite{Liang:2014tia}.  However,
although the peak height of the $f_{0} (980)$ is very similar, the
Breit--Wigner fit would provide larger branching fraction
$\mathcal{B}[D_{s}^{+} \to f_{0} (980) e^{+} \nu _{e}]$ than the
theoretical one.  Actually, integrating the theoretical mass
distribution in the range [$0.9 \gev$, $1.0 \gev$], we obtain the
branching fraction $\mathcal{B}[D_{s}^{+} \to f_{0} (980) e^{+} \nu
_{e}; \, f_{0} (980) \to \pi ^{+} \pi ^{-}] = 5.10 \times 10^{-4}$,
which is about four times smaller than the experimental value $2.00
\times 10^{-3}$.  Actually, in the experimental analysis of
Ref.~\cite{Ecklund:2009aa} different sources of background are
considered that make up for the lower mass region of the distribution.
The width of the $f_{0} (980)$ extracted in the analysis of
Ref.~\cite{Ecklund:2009aa} is $\Gamma = 91 ^{+30}_{-22} \pm 3 \mev$,
which is large compared to most experiments~\cite{Agashe:2014kda},
including the LHCb experiment of~\cite{Aaij:2014emv}, although the
admitted uncertainties are also large.  One should also take into
account that, while a Breit--Wigner distribution for the $f_{0} (980)$
is used in the analysis of Ref.~\cite{Ecklund:2009aa}, the large
coupling of the resonance to $K \bar{K}$ requires a Flatte form that
brings down fast the $\pi ^{+} \pi ^{-}$ mass distribution above the
$K \bar{K}$ threshold.  Our normalization in Fig.~\ref{fig:hist} is
done with the central value of the $\mathcal{B} [D^{+} \to (\pi ^{+}
K^{-})_{s\text{-wave}} e^{+} \nu _{e}]$ and no extra uncertainties
from this branching fraction are considered.  Yet, we find instructive
to do an exercise, adding to our results a ``background'' of $10
\text{ ns}^{-1} / \text{GeV}$ from different sources that our
calculation does not take into account, and then our signal for the
$f_{0} (980)$ has a good agreement with the peak of the experimental
distribution.

As mentioned above, the value extracted in~\cite{Ecklund:2009aa} for
the $f_{0} (980)$ signal is tied to the assumptions made, including
parts of the background that lead to a very large width of the
resonance, assuming a Breit--Wigner shape, etc.  Actually, in a more
recent paper~\cite{Hietala:2015jqa} the same CLEO data
of~\cite{Ecklund:2009aa} are reanalyzed taking a band of $f_{0} (980)$
masses within $60 \mev$ of $980 \mev$ and assuming a Flatte form of
the resonance and a rate for $\mathcal{B}[D_{s}^{+} \to f_{0} (980)
e^{+} \nu _{e}, \, f_{0} (980) \to \pi \pi ] = (0.13 \pm 0.02 \pm
0.01) \%$ is obtained.  This value is about a factor of two smaller
than the one reported in~\cite{Ecklund:2009aa} and more in agreement
with our results.

\begin{figure}[!t]
  \centering
  \Psfig{8.6cm}{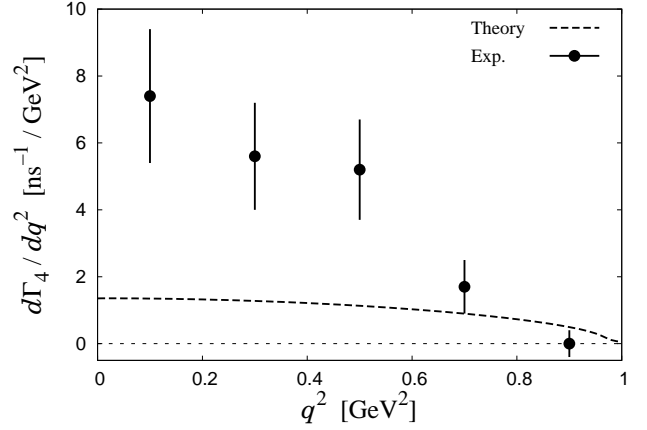}
  \caption{Differential decay width of the $D_{s}^{+} \to f_{0} (980)
    e^{+} \nu _{e}$ decay mode followed by $f_{0} (980) \to \pi ^{+}
    \pi ^{-}$, with $q^{2} = [M_{\rm inv}^{(l \nu )}]^{2}$.  The
    experimental data are taken from Ref.~\cite{Ecklund:2009aa}.  The
    experimental points should be rescaled dividing by about a factor
    of two if the absolute rate for the $f_{0} (980)$ production of
    the reanalysis of Ref.~\cite{Hietala:2015jqa} were used.  }
\label{fig:dG4dq2}
\end{figure}

Next we consider the differential decay width with respect to the
squared momentum transfer $q^{2}$, which coincides with the squared
invariant mass of the lepton pair: $q^{2} = [M_{\rm inv}^{( l \nu
  )}]^{2}$.  The differential decay width for the scalar meson
production is expressed as
\begin{align}
\frac{d \Gamma _{4}}{d q^{2}} 
= \frac{\left | G_{\rm F} \right | ^{2}}
{64 \pi ^{5} m_{D}^{3}}
& \int d M_{\rm inv}^{(h h)}
\frac{\left | V_{\rm had}^{(s)} \right | ^{2} P_{\rm cm}^{\prime} 
\tilde{p}_{h} \tilde{p}_{\nu} M_{\rm inv}^{( l \nu )}}{M_{\rm inv}^{(h h)}}
\notag \\
& \times 
\left ( \tilde{E}_{D} \tilde{E}_{S} - \frac{1}{3} | \tilde{\bm{p}}_{D} |^{2} 
\right ) .
\end{align}
This differential decay width was experimentally observed in
Ref.~\cite{Ecklund:2009aa} for the $D_{s}^{+} \to f_{0} (980) e^{+}
\nu _{e}$ decay mode followed by $f_{0} (980) \to \pi ^{+} \pi ^{-}$.
In this study we compare our theoretical value for this decay mode
with the experimental data in Fig.~\ref{fig:dG4dq2}.  The range of the
integral for $M_{\rm inv}^{(hh)}$ is [$0.9 \gev$, $1.0 \gev$].  As one
can see, we can to some extent reproduce the shape of the differential
decay width $d \Gamma _{4} / d q^{2}$ in experiment, but the absolute
value of the theoretical calculation is several times smaller than the
experimental one.  This can be, as we have explained, solved by
introducing background contributions when extracting the amount of the
$f_{0} (980)$ signal from experimental data.  Actually, as we have
commented before, the reanalysis of~\cite{Hietala:2015jqa} leads to
absolute values of the rate for the $f_{0} (980)$ production about a
factor of two smaller, and again if we scale the $q^{2}$ distribution
of in Fig.~\ref{fig:dG4dq2} by the factor the agreement is much
better.

\begin{table}[!b]
  \caption{Branching fractions of semileptonic $D$ decays into two
    pseudoscalar mesons in $s$ wave.  The branching fraction of 
    the $D^{+} \to (\pi ^{+} K^{-})_{s\text{-wave}} e^{+} \nu _{e}$ mode 
    is used as an input. }
  \label{tab:G4}  
  \begin{ruledtabular}
    \begin{tabular*}{8.6cm}{@{\extracolsep{\fill}}lccc}
      \multicolumn{4}{c}{$D_{s}^{+}$} 
      \\
      Mode & Range of $M_{\rm inv}^{(hh)}$ [GeV] & $l = e$ & $l = \mu$ \\
      \hline
      $\pi ^{+} \pi ^{-}$ & [0.9, 1.0] & 
      $5.10 \times 10^{-4}$  &  $4.71 \times 10^{-4}$ 
      \\
      $K^{+} K^{-}$ & [$2 m_{K}$, 1.2] & 
      $1.42 \times 10^{-4}$ & $1.30 \times 10^{-4}$ 
      \\
      $\pi ^{-} K^{+}$ & [$m_{\pi} + m_{K}$, 1.0] & 
      $8.11 \times 10^{-5}$ & $7.63 \times 10^{-5}$ 
      \\
      \\
      \multicolumn{4}{c}{$D^{+}$} 
      \\
      Mode & Range of $M_{\rm inv}^{(hh)}$ [GeV] & $l = e$ & $l = \mu$ \\
      \hline
      $\pi ^{+} \pi ^{-}$ & [$2 m_{\pi}$, 1.0] & 
      $5.11 \times 10^{-4}$ & $4.85 \times 10^{-4}$ 
      \\
      $\pi ^{0} \eta$ & [$m_{\pi} + m_{\eta}$, 1.1] & 
      $6.37 \times 10^{-5}$ & $5.86 \times 10^{-5}$ 
      \\
      $K^{+} K^{-}$ & [$2 m_{K}$, 1.2] & 
      $2.24 \times 10^{-6}$ & $2.01 \times 10^{-6}$ 
      \\
      $\pi ^{+} K^{-}$ & [$m_{\pi} + m_{K}$, 1.0] & 
      $2.32 \times 10^{-3}$ & $2.16 \times 10^{-3}$ 
      \\
      \\
      \multicolumn{4}{c}{$D^{0}$} 
      \\
      Mode & Range of $M_{\rm inv}^{(hh)}$ [GeV] & $l = e$ & $l = \mu$ \\
      \hline
      $\pi ^{-} \eta$ & [$m_{\pi} + m_{\eta}$, 1.1] & 
      $4.93 \times 10^{-5}$ & $4.53 \times 10^{-5}$ 
      \\
      $K^{0} K^{-}$ & [$2 m_{K}$, 1.2] & 
      $5.47 \times 10^{-6}$ & $4.88 \times 10^{-6}$ 
      \\
      $\pi ^{-} \bar{K}^{0}$ & [$m_{\pi} + m_{K}$, 1.0] & 
      $8.99 \times 10^{-4}$ & $8.38 \times 10^{-4}$ 
      \\
    \end{tabular*}
  \end{ruledtabular}
\end{table}

Moreover, integrating the mass distributions we calculate the
branching fractions of the semileptonic $D$ mesons into two
pseudoscalar mesons in $s$ wave, which are listed in
Table~\ref{tab:G4}.  We note that the branching fraction
$\mathcal{B}[D^{+} \to (\pi ^{+} K^{-})_{s\text{-wave}} e^{+} \nu
_{e}]= 2.32 \times 10^{-3}$ is used as an input to fix the common
constant, $C = 4.597$.  Among the listed values, we can compare the
theoretical and experimental values of the branching fraction
$\mathcal{B}[D_{s}^{+} \to (K^{+} K^{-})_{s\text{-wave}} e^{+} \nu
_{e}]$.  Namely, in Ref.~\cite{Aubert:2008rs} this branching fraction
is obtained as $( 0.22 ^{+0.12}_{-0.08} \pm 0.03) \%$ of the total
$D_{s}^{+} \to K^{+} K^{-} e^{+} \nu _{e}$, which is dominated by the
$\phi (1020)$ vector meson.  This indicates, together with the
branching fraction $D_{s}^{+} \to \phi (1020) e^{+} \nu _{e}$, we can
estimate $\mathcal{B}[D_{s}^{+} \to (K^{+} K^{-})_{s\text{-wave}}
e^{+} \nu _{e}] = (5.5^{+3.1}_{-2.1}) \times 10^{-5}$.  Theoretically
this is $1.42 \times 10^{-4}$.  Although our value overestimates the
mean value of the experimental data, it is still in $3 \sigma$ errors
of the experimental value.

\subsection{Production of vector mesons}
\label{sec:vector}

\begin{table}
  \caption{Branching fractions of semileptonic $D$ decays into vector
    mesons. }
  \label{tab:G3}  
  \begin{ruledtabular}
    \begin{tabular*}{8.6cm}{@{\extracolsep{\fill}}lcc}
      \multicolumn{3}{c}{$D_{s}^{+}$} 
      \\
      Mode & $l = e$ & $l = \mu$ \\
      \hline
      $\phi (1020)$ & 
      $2.12 \times 10^{-2}$  &  $1.94 \times 10^{-2}$ 
      \\
      $K^{\ast} (892)^{0}$ & 
      $2.02 \times 10^{-3}$ & $1.89 \times 10^{-3}$ 
      \\
      \\
      \multicolumn{3}{c}{$D^{+}$} 
      \\
      Mode & $l = e$ & $l = \mu$ \\
      \hline
      $\bar{K}^{\ast} (892)^{0}$ & 
      $5.56 \times 10^{-2}$ & $5.12 \times 10^{-2}$ 
      \\
      $\rho (770)^{0}$ & 
      $2.54 \times 10^{-3}$ & $2.37 \times 10^{-3}$ 
      \\
      $\omega (782)$ & 
      $2.46 \times 10^{-3}$ & $2.29 \times 10^{-3}$ 
      \\
      \\
      \multicolumn{3}{c}{$D^{0}$} 
      \\
      Mode & $l = e$ & $l = \mu$ \\
      \hline
      $K^{\ast} (892)^{-}$ & 
      $2.15 \times 10^{-2}$ & $1.98 \times 10^{-2}$ 
      \\
      $\rho (770)^{-}$ & 
      $1.97 \times 10^{-3}$ & $1.84 \times 10^{-3}$ 
      \\
    \end{tabular*}
  \end{ruledtabular}
\end{table}

\begin{figure}[!t]
  \centering
  \Psfig{8.6cm}{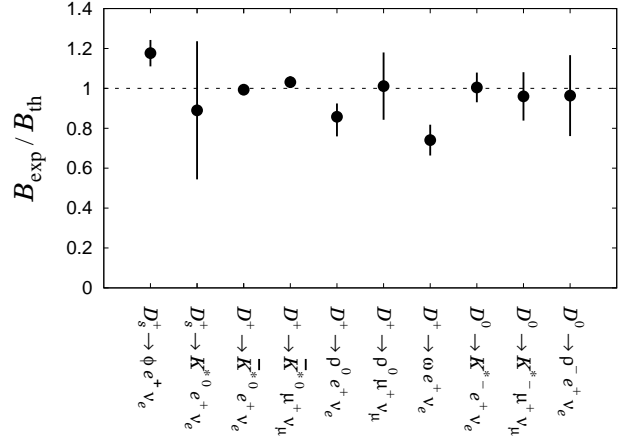}
  \caption{Ratio of the experimental to theoretical branching
    fractions for the semileptonic $D$ decays into vector mesons. }
\label{fig:vec}
\end{figure}

Let us move to the vector meson productions in the semileptonic $D$
decays.  For the vector mesons we fix the common prefactor
$C^{\prime}$ so as to reproduce the 10 available experimental
branching fractions listed in Table~\ref{tab:Br}.  From the best fit
we obtain the value $C^{\prime} = 1.563 \gev$, which gives $\chi ^{2}
/ N_{\rm d.o.f.} = 22.8 / 9 \approx 2.53$.  The theoretical values of
the branching fractions are listed in Table~\ref{tab:G3} and are
compared with the experimental data in Fig.~\ref{fig:vec}, where we
plot the ratio of the experimental to theoretical branching fractions.
We calculate the experimental branching fraction of the $D^{+} \to
\bar{K} (892)^{0} l^{+} \nu _{l}$ ($l = e$ and $\mu$) process by
dividing the value in Table~\ref{tab:Br} by the branching fraction
$\mathcal{B}[\bar{K}^{\ast} (892)^{0} \to K^{-} \pi ^{+}] = 2/3$,
which is obtained with isospin symmetry.  As one can see from
Fig.~\ref{fig:vec}, the experimental values are reproduced well solely
by the model parameter $C^{\prime}$ with $\chi ^{2} / N_{\rm d.o.f.}
\approx 2.53$.

\begin{figure}[!b]
  \centering
  \Psfig{8.6cm}{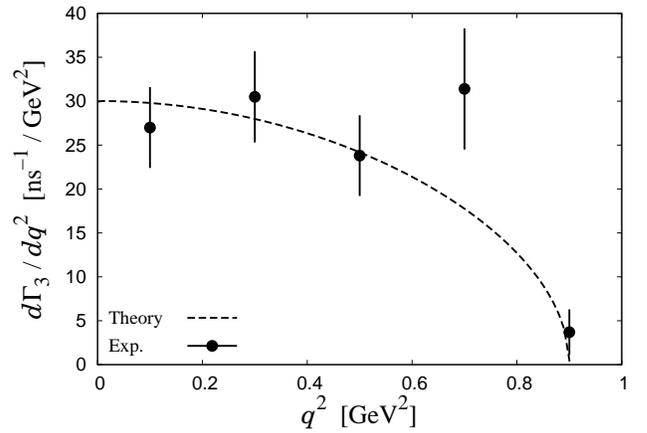}
  \caption{Differential decay width of the $D_{s}^{+} \to \phi (1020)
    e^{+} \nu _{e}$ decay mode followed by $\phi (1020) \to K^{+}
    K^{-}$, with $q^{2} = [M_{\rm inv}^{(l \nu )}]^{2}$.  The
    experimental data are taken from Ref.~\cite{Ecklund:2009aa}.  The
    theoretical value is multiplied by the branching fraction of the
    $\phi (1020)$ meson to $K^{+} K^{-}$, $\mathcal{B} [\phi (1020)
    \to K^{+} K^{-}] = 48.9 \%$~\cite{Agashe:2014kda}. }
\label{fig:dG3dq2}
\end{figure}

Next for the $D_{s}^{+} \to \phi (1020) e^{+} \nu _{e}$ decay mode we
consider the differential decay width with respect to the squared
momentum transfer $q^{2}$, which coincides with the squared invariant
mass of the lepton pair: $q^{2} = [M_{\rm inv}^{( l \nu )}]^{2}$.
This differential decay width was already measured in an
experiment~\cite{Ecklund:2009aa} for the $D_{s}^{+} \to \phi (1020)
e^{+} \nu _{e}$ decay mode.  In a similar manner to the previous case,
the differential decay width for the vector meson production is
expressed as
\begin{equation}
\frac{d \Gamma _{3}}{d q^{2}}
= \frac{\left | G_{\rm F} V_{\rm had}^{(v)} \right | ^{2}}
{16 \pi ^{3} m_{D}^{3} m_{V}}
P_{\rm cm} \tilde{p}_{\nu} M_{\rm inv}^{(l \nu )} 
\left ( \tilde{E}_{D} \tilde{E}_{V} - \frac{1}{3} | \tilde{\bm{p}}_{D} |^{2} 
\right ) .
\end{equation}
In Fig.~\ref{fig:dG3dq2} we compare our result for this reaction with
the experimental one.  As one can see, our theoretical result
reproduces the experimental value of the differential decay width
quantitatively well.  This means that our method to calculate the
semileptonic decays of $D$ mesons is good enough to describe the
decays into vector mesons.

In this study we have not evaluated the $D^{+} \to \phi (1020) e^{+}
\nu _{e}$ decay. This decay proceeds like the $D^{+} \to \omega (782)
e^{+} \nu _{e}$ decay that we have evaluated and one has a $d \bar{d}$
at the end.  Since the $\phi$ is $s \bar{s}$ then this is forbidden in
our approach, at the tree level that we have considered for the vector
production.  Experimentally, this rate is $< 9 \times 10^{-5}$. This
is an upper bound about $30$ times smaller than the rate of the omega
production that we have evaluated. We do not want to go beyond, but
can give some idea on how a finite rate could be obtained in our
approach. For this one would have to hadronize the $d \bar{d}$ into a
$K^{0} \bar{K}^{0}$, then have a loop for $K^{0} \bar{K}^{0}$
propagation in $p$-wave and finally have the $K^{0} \bar{K}^{0}$
couple to the $\phi$.  Some technical details could be borrowed from
the study of $\phi \to \pi \pi$ decay studied in~\cite{Oller:1999ag}
but one can get an indication that the rate should be rather small by
simply noting that the hadronization to meson--meson pairs has a
reduction factor, as one can see by comparing for instance $f_{0}
(500)$ production with $\rho$ production~\cite{Bayar:2014qha}.  On the
other hand, the coupling of $\phi$ to $K^{0} \bar{K}^{0}$ is
intrinsically small, as one can see from the $1.5 \mev$ partial decay
width of this channel [comparatively the $\Delta (1232)$ partial decay
width to the $\pi N$ channel would be about $15 \mev$ for a pion with
the same momentum as the kaon in the $\phi$ decay].  There are other
factors to consider, but this can give us a feeling that the rate
could be some orders of magnitude smaller than for omega production.

\subsection{Comparison between scalar and vector meson contributions}
\label{sec:compare}

Finally we compare the mass distributions of the two pseudoscalar
mesons in $s$- and $p$-wave contributions.  In the present approach
the $s$-wave part comes from the rescattering of two pseudoscalar
mesons including the scalar meson contribution, while the $p$-wave one
from the decay of a vector meson.  In this study we consider three
decay modes: $D_{s}^{+} \to \pi ^{+} \pi ^{-} e^{+} \nu _{e}$,
$D_{s}^{+} \to K^{+} K^{-} e^{+} \nu _{e}$, and $D^{+} \to \pi ^{+}
K^{-} e^{+} \nu _{e}$.  The $D^{+} \to \pi ^{+} \pi ^{-} e^{+} \nu
_{e}$ decay mode would have a large $p$-wave contribution from $\rho
(770)$, but we do not consider this decay mode since it is a Cabibbo
suppressed process.

First we consider the $D_{s}^{+} \to \pi ^{+} \pi ^{-} e^{+} \nu _{e}$
decay mode.  This is a specially clean mode, since it does not have
vector meson contributions and is dominated by the $s$-wave part.
Namely, while the $\pi ^{+} \pi ^{-}$ can come from a scalar meson,
the primary quark--antiquark pair in the semileptonic $D_{s}^{+}$
decay is $s \bar{s}$, which is isospin $I = 0$ and hence the $\rho
(770)$ cannot contribute to the $\pi ^{+} \pi ^{-}$ mass distribution.
The primary $s \bar{s}$ can be $\phi (1020)$, but it decays dominantly
to $K \bar{K}$ and the $\phi (1020) \to \pi ^{+} \pi ^{-}$ decay is
negligible.  This fact enables us to observe the scalar meson peak
without a contamination from vector meson decays and discuss the quark
configuration in the $f_{0} (980)$ resonance as in
Sec.~\ref{sec:scalar}.

\begin{figure}[!t]
  \centering
  \Psfig{8.6cm}{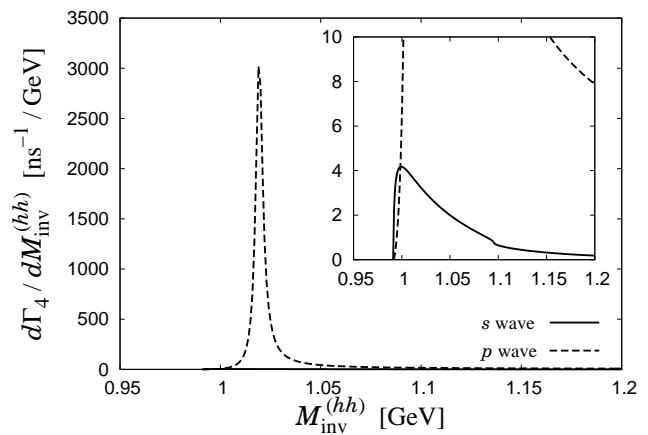}
  \caption{$K^{+} K^{-}$ invariant mass distribution for the
    semileptonic decay $D_{s}^{+} \to K^{+} K^{-} e^{+} \nu _{e}$ both
    in $s$ and $p$ waves.  }
\label{fig:KpKm}
\end{figure}

Next let us consider the $D_{s}^{+} \to K^{+} K^{-} e^{+} \nu _{e}$
decay mode.  As we have seen, the $K^{+} K^{-}$ mass distribution in
$s$ wave is a consequence of the $f_{0} (980)$ tail.  However, its
contribution should be largely contaminated by the $\phi (1020) \to
K^{+} K^{-}$ in $p$ wave, which has a larger branching fraction than
the $( K^{+} K^{-} )_{s\text{-wave}}$ in the semileptonic decay.  In
order to see this, we calculate the $p$-wave $K^{+} K^{-}$ mass
distribution for $D_{s}^{+} \to K^{+} K^{-} e^{+} \nu _{e}$, which can
be formulated as
\begin{equation}
\frac{d \Gamma_{3}}{d M_{\rm inv}^{(hh)}} 
= - \frac{2 m_{V}}{\pi} \text{Im }
\frac{\Gamma_{3} \times \mathcal{B}[V \to h h]}{
[ M_{\rm inv}^{(hh)} ]^{2} - m_{V}^{2} 
+ i m_{V} \Gamma _{V} ( M_{\rm inv}^{(hh)} ) } ,
\end{equation}
where $m_{V}$ is the vector meson mass and the energy dependent decay
width $\Gamma _{V} ( M_{\rm inv}^{(hh)})$ is defined as
\begin{equation}
\Gamma _{V} ( M_{\rm inv}^{(hh)}) \equiv 
\bar{\Gamma}_{V} 
\left ( \frac{p^{\rm off} ( M_{\rm inv}^{(hh)})}{p^{\rm on}} \right )^{3} , 
\end{equation}
\begin{equation}
p^{\rm off} ( M_{\rm inv}^{(hh)}) \equiv 
\frac{\lambda ^{1/2} 
( [ M_{\rm inv}^{(hh)}]^{2}, \, m_{h}^{2}, \, m_{h}^{\prime 2})}
{2 M_{\rm inv}^{(hh)}} , 
\end{equation}
\begin{equation}
p^{\rm on} \equiv 
\frac{\lambda ^{1/2} ( m_{V}^{2}, \, m_{h}^{2}, \, m_{h}^{\prime 2})}
{2 m_{V}} .
\end{equation}
For the $\phi (1020)$ meson we take $\bar{\Gamma}_{\phi} = 4.27 \mev$
and $\mathcal{B}[\phi \to K^{+} K^{-}] = 0.489$~\cite{Agashe:2014kda}.
The numerical result for the $(K^{+} K^{-})_{p\text{-wave}}$ mass
distribution is shown in Fig.~\ref{fig:KpKm} together with the $(K^{+}
K^{-})_{s\text{-wave}}$.  From the figure, compared to the $(K^{+}
K^{-})_{p\text{-wave}}$ contribution we cannot find any significant
$(K^{+} K^{-})_{s\text{-wave}}$ contribution, which was already noted
in the experimental mass distribution in Ref.~\cite{Aubert:2008rs}.
Nevertheless, we emphasize that the $(K^{+} K^{-})_{s\text{-wave}}$
fraction of the semileptonic $D_{s}^{+}$ decay is large enough to be
extracted~\cite{Aubert:2008rs}.  Actually in Ref.~\cite{Aubert:2008rs}
they extracted the $(K^{+} K^{-})_{s\text{-wave}}$ fraction by
analysing the interference between the $s$- and $p$-wave
contributions.  This fact, and the qualitative reproduction of the
branching fractions in our model, implies that the $f_{0} (980)$
resonance couples to the $K \bar{K}$ channel with a certain strength,
which can be translated into the $K \bar{K}$ component in $f_{0}
(980)$, in a similar manner to the $K D$ component in $D_{s 0}^{\ast}
(2317)$~\cite{Albaladejo:2015kea, Navarra:2015iea}, in terms of the
compositeness~\cite{Sekihara:2014kya}.  Anyway, in order to conclude
the structure of the $f_{0} (980)$ more clearly, it is important to
reduce the experimental errors on the $(K^{+} K^{-})_{s\text{-wave}}$.

\begin{figure}[!t]
  \centering
  \Psfig{8.6cm}{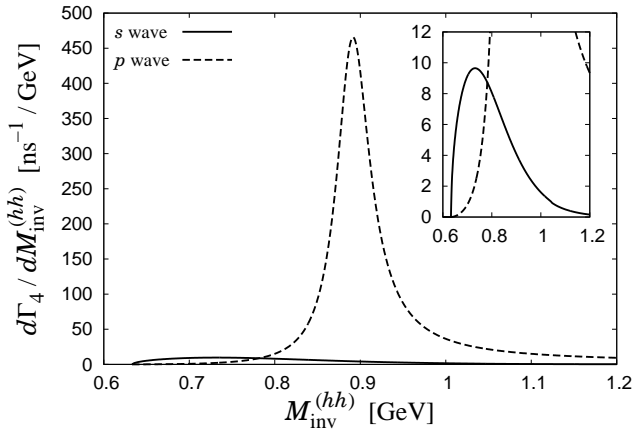}
  \caption{$\pi ^{+} K^{-}$ invariant mass distribution for the
    semileptonic decay $D^{+} \to \pi ^{+} K^{-} e^{+} \nu _{e}$ both
    in $s$ and $p$ waves.  }
\label{fig:ppKm}
\end{figure}

Finally we consider the $D^{+} \to \pi ^{+} K^{-} e^{+} \nu _{e}$
decay mode.  In this mode the $(\pi ^{+} K^{-})_{s\text{-wave}}$ from
the $K_{0}^{\ast}(800)$ and the $(\pi ^{+} K^{-})_{p\text{-wave}}$
from the $K^{\ast}(892)$ are competing with each other.  In a similar
manner to the $D_{s}^{+} \to K^{+} K^{-} e^{+} \nu _{e}$ case, we
calculate the mass distribution also for the $p$-wave $\pi ^{+} K^{-}$
contribution $d \Gamma _{3} / d M_{\rm inv}^{(hh)}$ with $\bar{\Gamma}
_{K^{\ast}} = 49.1 \mev$~\cite{Agashe:2014kda}, and the result is
shown in Fig.~\ref{fig:ppKm}.  As one can see, thanks to the width of
$\sim 50 \mev$ for the $K^{\ast} (892)$, the $s$-wave component can
dominate the mass distribution below $0.8 \gev$.  We note that we
would obtain an almost similar result for the $D^{0} \to \pi ^{-}
\bar{K}^{0} e^{+} \nu _{e}$ decay mode due to isospin symmetry.

As to the theoretical uncertainties, we can play a bit with the
cut-offs used to regularize the loops, such that the masses of the
states do not change appreciably.  This exercise has been done a
number of times and given us the feeling that within our models the
uncertainties are below $10 \%$.  For the case of scalar production
where we have a range of invariant masses and rely upon a constant
production vertex $C$, the changes with the invariant mass in the
primary form factors, prior to the final state interaction of the
mesons, as found in~\cite{Kang:2013jaa}, would add some extra
uncertainty. In total it would be fair to accept about $20 \%$
uncertainties in this case in the limited range of energies that we
move.

\section{Conclusion}
\label{sec:conclusion}

In this study we have discussed the semileptonic decays of $D$ mesons
into light scalar and vector mesons.  For the scalar meson production,
we have formulated the semileptonic decay as the combination of two
parts.  One is the weak decay of the charm quark and the emission
of a lepton pair via the $W$ boson.  The other is a simple
hadronization of light $q \bar{q}$ pair plus an extra $\bar{q} q$ from
vacuum into two pseudoscalar mesons after the $W$ boson emission, so
as to generate the scalar mesons dynamically in the meson--meson final
state interaction.  The hadronization naturally gives the weight of
each pair of pseudoscalar mesons in the decay process, which governs
which scalar meson appears in the decay mode.  For the vector mesons,
on the other hand, we have not considered the hadronization with an
extra $\bar{q} q$ and have directly used the light $q \bar{q}$ pair
after the $W$ boson emission as a weight for the vector mesons, which
are expected to be genuinely $q \bar{q}$ states.  We note that we can
specify flavors of quarks contained in the final state scalar and vector
mesons by considering Cabibbo favored and suppressed processes.  In
addition, since the leptons interact only weakly, the semileptonic
decay of the heavy meson to two light mesons $+ l^{+} \nu _{l}$ brings
a suitable condition to measure effects of the final state interaction
of the two light mesons.

In our model of the semileptonic decay, the production yields of the
scalar and vector mesons are respectively determined solely by
constant prefactors $C$ and $C^{\prime}$ as model parameters.  Fixing
$C$ from the branching fraction of the $D^{+} \to ( \pi ^{+} K^{-}
)_{s\text{-wave}} e^{+} \nu _{e}$ decay, we have calculated branching
fractions of scalar meson productions.  We have qualitatively
reproduced the experimental value of the branching fractions of
$D_{s}^{+} \to ( \pi ^{+} \pi ^{-} )_{s\text{-wave}} e^{+} \nu _{e}$
and $( K^{+} K^{-} )_{s\text{-wave}} e^{+} \nu _{e}$ decay modes.
Some deviations of these branching fractions compared to the
experimental values can be explained by taking into account the
background of the mass distribution for the $\pi ^{+} \pi ^{-}$ case
and by the large experimental error for the $K^{+} K^{-}$ case.  For
the vector mesons, we have determined the constant $C^{\prime}$ so as
to fit our numerical values to the available experimental values of
the branching fractions, and we have reproduced the experimental
values at a quantitative level.  We also compared the mass
distributions of the two pseudoscalar mesons in $s$- and $p$-wave
contributions, which come from decays of the scalar and vector mesons,
respectively.

We have found that the Cabibbo favored decay mode $D_{s}^{+} \to f_{0}
(980) l^{+} \nu _{l}$ followed by $f_{0} (980) \to \pi ^{+} \pi ^{-}$
and $K^{+} K^{-}$ is of special interest.  For the $f_{0} (980) \to
\pi ^{+} \pi ^{-}$ mode, we have found that there is no $p$-wave
contamination from $\rho (770)$ decay and hence it should be dominated
by the $s$-wave part.  Then, we have confirmed the experimental fact
that the $\pi ^{+} \pi ^{-}$ mass distribution shows a clear $f_{0}
(980)$ signal while the $f_{0} (500)$ contribution is negligible.
This strongly indicates that the $f_{0} (980)$ has a substantial
fraction of the strange quarks while the $f_{0} (500)$ has a
negligible strange quark component.  For the $f_{0} (980) \to K^{+}
K^{-}$ mode, on the other hand, the $(K^{+} K^{-})_{s\text{-wave}}$
contribution is highly contaminated by the $\phi (1020) \to K^{+}
K^{-}$ decay in $p$ wave.  Nevertheless, the $(K^{+}
K^{-})_{s\text{-wave}}$ fraction of the semileptonic $D_{s}^{+}$ decay
is large enough to be extracted experimentally, which implies that the
$f_{0} (980)$ resonance couples to the $K \bar{K}$ channel with a
certain strength and hence implies a certain amount of the $K \bar{K}$
component in $f_{0} (980)$.

\begin{acknowledgments}
  We appreciate information from S.~Stone and X.~W.~Kang.
  We acknowledge the support by Open Partnership Joint Projects of
  JSPS Bilateral Joint Research Projects.  This work is partly
  supported by Grants-in-Aid for Scientific Research from MEXT and
  JSPS (No.~15K17649 and No.~15J06538), the Spanish Ministerio de
  Economia y Competitividad and European FEDER funds under the
  contract number FIS2011-28853-C02-01 and FIS2011-28853-C02-02, and
  the Generalitat Valenciana in the program Prometeo II-2014/068.  We
  acknowledge the support of the European Community-Research
  Infrastructure Integrating Activity Study of Strongly Interacting
  Matter (acronym HadronPhysics3, Grant Agreement n.~283286) under the
  Seventh Framework Program of the EU.
  We are deeply grateful to the Yukawa Institute for Theoretical
  Physics, Kyoto University, where this work was initiated during the
  YITP workshop YITP-T-14-03 on ``Hadrons and hadron interactions in
  QCD''.

\end{acknowledgments}

\appendix 

\section{Conventions}
\label{app:1}

In this Appendix we summarize conventions used in this study.  

\subsection{Metric and Lorentz indices}

In this article the metric in four-dimensional Minkowski space is
$g^{\mu \nu} = g_{\mu \nu} = \text{diag}(1, \, -1, \, -1, \, -1)$ and
the Einstein summation convention is used unless explicitly mentioned.
The scalar product of two vectors $a^{\mu}$ and $b^{\mu}$ is
represented as $a \cdot b = a_{\mu} b^{\mu} = a^{0} b^{0} - \bm{a}
\cdot \bm{b}$.

\subsection{Dirac spinors and matrices}

As the positive and negative energy solutions of the Dirac equation,
we express the Dirac spinors respectively as $u ( \bm{p}, \, s)$ and
$v ( \bm{p}, \, s)$, where $\bm{p}$ is three-momentum of the field and
$s$ represents its spin.  The Dirac spinors are normalized as follows:
\begin{equation}
\overline{u}(\bm{p}, \, s) u(\bm{p}, \, s^{\prime}) 
= \delta _{s s^{\prime}} , 
\quad 
\overline{v}(\bm{p}, \, s) v(\bm{p}, \, s^{\prime}) 
= - \delta _{s s^{\prime}} , 
\end{equation}
with $\overline{u} \equiv u^{\dagger} \gamma ^{0}$ and $\overline{v}
\equiv v^{\dagger} \gamma ^{0}$, and hence we have
\begin{equation}
\begin{split}
& \sum _{s} 
u(\bm{p}, \, s) \overline{u}(\bm{p}, \, s) 
= \frac{\Slash{p} + m}{2 m} , 
\\
& \sum _{s} 
v(\bm{p}, \, s) \overline{v}(\bm{p}, \, s) 
= \frac{\Slash{p} - m}{2 m} , 
\end{split}
\end{equation}
where $m$ is the mass of the field, $\Slash{p} \equiv \gamma ^{\mu}
p_{\mu}$ with $\gamma ^{\mu}$ being the Dirac $\gamma$ matrices, and
$p^{\mu} \equiv \left ( \sqrt{\bm{p}^{2} + m^{2}}, \, \bm{p} \right )$
is the on-shell four-momentum of the solution.

The identities for the Dirac matrices used in this study are
summarized as follows:
\begin{equation}
\gamma ^{0} ( \gamma ^{\mu} )^{\dagger} \gamma ^{0} = \gamma ^{\mu} , 
\quad 
( \gamma _{5} )^{\dagger} = \gamma _{5} , 
\end{equation}
\begin{equation}
\trace \left [ \gamma ^{\mu} \gamma ^{\nu} 
\gamma ^{\rho} \gamma ^{\sigma} \right ]
= 4 ( g^{\mu \nu} g^{\rho \sigma} - g^{\mu \rho} g^{\nu \sigma} 
+ g^{\mu \sigma} g^{\nu \rho} ) ,
\end{equation}
\begin{equation}
\trace \left [ \gamma _{5} \gamma ^{\mu} \gamma ^{\nu} 
\gamma ^{\rho} \gamma ^{\sigma} \right ]
= - 4 i \epsilon ^{\mu \nu \rho \sigma} ,
\end{equation}
\begin{equation}
\trace \left [ \gamma  ^{\mu} \gamma ^{\nu} 
\gamma ^{\rho} \right ] 
= \trace \left [ \gamma _{5} \gamma  ^{\mu} \gamma ^{\nu} 
\gamma ^{\rho} \right ] 
= 0 , 
\end{equation}
where $\gamma _{5} \equiv i \gamma ^{0} \gamma ^{1} \gamma ^{2} \gamma
^{3}$ and $\epsilon ^{\mu \nu \rho \sigma}$ is the Levi-Civita symbol
with the normalization $\epsilon ^{0 1 2 3} = 1$.  The Levi-Civita
symbol satisfies the following identity
\begin{equation}
\epsilon ^{\alpha \beta \mu \nu} \epsilon _{\alpha \beta \rho \sigma}
= - 2 ( g^{\mu}_{\rho} g^{\nu}_{\sigma} - g^{\mu}_{\sigma} g^{\nu}_{\rho} ) .
\end{equation}

\subsection{Isospin basis}

In terms of the isospin states $| I , \, I_{3} \rangle$, the phase
convention for pseudoscalar mesons is given by
\begin{equation}
| \pi ^{+} \rangle = - | 1, \, 1 \rangle , 
\quad 
| K^{-} \rangle 
= - | 1/2, \, - 1/2 \rangle , 
\end{equation}
while other pseudoscalar mesons are represented without phase factors.
As a result, we can translate the physical two-pseudoscalar meson
states into the isospin basis, which we specify as $(I, \, I_{3})$, as
\begin{equation}
| K \bar{K} (0, \, 0) \rangle = 
- \frac{1}{\sqrt{2}} | K^{+} K^{-} \rangle 
- \frac{1}{\sqrt{2}} | K^{0} \bar{K}^{0} \rangle , 
\end{equation}
\begin{equation}
| \eta \eta (0 , \, 0) \rangle = 
| \eta \eta \rangle ,
\end{equation}
\begin{equation}
| K \bar{K} (1, \, 0) \rangle = 
- \frac{1}{\sqrt{2}} | K^{+} K^{-} \rangle 
+ \frac{1}{\sqrt{2}} | K^{0} \bar{K}^{0} \rangle , 
\end{equation}
\begin{equation}
| K \bar{K} (1, \, -1) \rangle = 
- | K^{0} K^{-} \rangle ,
\end{equation}
\begin{equation}
| \pi \eta (1, \, 0) \rangle = 
| \pi ^{0} \eta \rangle , 
\end{equation}
\begin{equation}
| \pi \eta (1, \, -1) \rangle = 
| \pi ^{-} \eta \rangle , 
\end{equation}
\begin{equation}
| \pi K (1/2, \, -1/2) \rangle = 
\frac{1}{\sqrt{3}} | \pi ^{0} K^{0}\rangle 
- \sqrt{\frac{2}{3}} | \pi ^{-} K^{+} \rangle , 
\end{equation}
\begin{equation}
| \pi \bar{K} (1/2, \, 1/2) \rangle = 
\sqrt{\frac{2}{3}} | \pi ^{+} K^{-} \rangle 
- \frac{1}{\sqrt{3}} | \pi ^{0} \bar{K}^{0} \rangle , 
\end{equation}
\begin{equation}
| \pi \bar{K} (1/2, \, -1/2) \rangle = 
- \frac{1}{\sqrt{3}} | \pi ^{0} K^{-} \rangle 
- \sqrt{\frac{2}{3}} | \pi ^{-} \bar{K}^{0} \rangle , 
\end{equation}

Furthermore, the vector meson states are represented in terms of
quarks as
\begin{equation}
| \rho ^{0} \rangle = \frac{1}{\sqrt{2}} | u \bar{u} \rangle 
- \frac{1}{\sqrt{2}} | d \bar{d} \rangle , 
\quad 
| \rho ^{-} \rangle = | d \bar{u} \rangle ,
\end{equation}
\begin{equation}
| \omega \rangle = \frac{1}{\sqrt{2}} | u \bar{u} \rangle 
+ \frac{1}{\sqrt{2}} | d \bar{d} \rangle , 
\end{equation}
\begin{equation}
| K^{\ast 0} \rangle = | d \bar{s} \rangle , 
\quad 
| \bar{K}^{\ast 0} \rangle = | s \bar{d} \rangle , 
\quad 
| \bar{K}^{\ast -} \rangle = - | s \bar{u} \rangle . 
\end{equation}

\subsection{Feynman rules}

The $W \nu l$ coupling is expressed as
\begin{equation}
- i V_{W \nu l}^{\mu} = i \frac{g_{\rm W}}{\sqrt{2}} \gamma ^{\mu}
\frac{1 - \gamma _{5}}{2} ,
\end{equation}
with $g_{\rm W}$ being the coupling constant of the weak interaction,
and the $W c q$ coupling as
\begin{equation}
- i V_{W c q}^{\mu} = i \frac{g_{\rm W} V_{c q}}{\sqrt{2}} \gamma ^{\mu}
\frac{1 - \gamma _{5}}{2} ,
\end{equation}
where $V_{c q}$ is the Cabibbo--Kobayashi--Maskawa matrix elements for
the transition from the charm to light quark $q$.  The $W$ boson
propagator with four-momentum $p^{\mu}$ is written as
\begin{equation}
i P_{W}^{\mu \nu} ( p ) =  \frac{- i g^{\mu \nu}}{p^{2} - M_{W}^{2} + i 0} ,
\end{equation}
with the mass of the $W$ boson $M_{W}$.  The coupling constant $g_{\rm
  W}$ and the mass of the $W$ boson $M_{W}$ are related to the Fermi
coupling constant $G_{\rm F}$ as
\begin{equation}
G_{\rm F} = \frac{g_{\rm W}^{2}}{4 \sqrt{2} M_{W}^{2}} .
\end{equation}

\subsection{Physical constants}

In this article we use the following values for physical constants.
The Fermi coupling constant: $G_{\rm F} \approx 1.166 \times 10^{-5}
\gev ^{-2}$.  The Cabibbo--Kobayashi--Maskawa matrix elements: $| V_{c
  s} | \approx 0.986$ and $| V_{c d} | \approx 0.225$.  The masses of
heavy mesons: $m_{D_{s}^{+}} = 1968.30 \mev$, $m_{D^{+}} = 1869.61
\mev$, and $m_{D^{0}} = 1864.84 \mev$.  Isospin symmetric masses are
used for the light mesons: $m_{\pi} = 138.04 \mev$, $m_{K} = 495.67
\mev$, and $m_{\eta} = 547.85 \mev$ for the pseudoscalar mesons, and
$m_{\rho} = 775.19 \mev$, $m_{\omega} = 782.65 \mev$, $m_{K^{\ast}} =
893.74 \mev$, and $m_{\phi} = 1019.46 \mev$ for the vector mesons.
The masses of the leptons: $m_{e} = 0.511 \mev$, $m_{\mu} = 105.66
\mev$, and $m_{\nu _{e}} = m_{\nu _{\mu}} = 0 \mev$.

\end{document}